\documentclass[aps,10pt,prb,superscriptaddress,preprintnumbers,twocolumn,floatfix,nofootinbib]{revtex4-2}
\pdfoutput=1
\usepackage{mathtools}
\usepackage{amsthm}
\usepackage{amsfonts,amssymb }
\usepackage{graphicx} 
\usepackage{physics}
\usepackage{xcolor}
\usepackage{enumitem}
\usepackage{dsfont}
\usepackage{bm}
\usepackage[hidelinks]{hyperref}

\begin{document}
\newcommand{\E}[0]{\mathop{{}\mathbb{E}}}
\newcommand{\Pro}[0]{\mathop{{}\mathbb{P}}}

\title{Quantum thermalization must occur in translation-invariant systems \\ at high temperature}
\author{Sa\'ul Pilatowsky-Cameo}
\email{saulpila@mit.edu}
\affiliation{Center for Theoretical Physics, Massachusetts Institute of Technology, Cambridge, MA 02139, USA}

\author{Soonwon Choi}
\email{soonwon@mit.edu}
\affiliation{Center for Theoretical Physics, Massachusetts Institute of Technology, Cambridge, MA 02139, USA}

 \newtheorem{theorem}{Theorem}
    \newtheorem{corollary}{Corollary}
    \newtheorem{lemma}{Lemma}
    \newtheorem{prop}{Proposition}

\theoremstyle{definition}
  \newtheorem{definition}{Definition}

\begin{abstract}
 Quantum thermalization describes how closed quantum systems can effectively reach thermal equilibrium, resolving the apparent
incongruity between the reversibility of Schr\"odinger's equation and the second law of thermodynamics.
Despite its ubiquity and conceptual significance, the precise conditions that give rise to quantum thermalization are still not well understood. After nearly a century of efforts, we have yet to find a complete mathematical proof that an effective statistical description naturally emerges the underlying quantum dynamics in generic settings. Here, we prove that quantum thermalization must occur in any qubit system with local interactions under three conditions: (i)~high effective temperature, (ii)~translation invariance, and (iii)~no perfect resonances in the energy spectrum.
Specifically, we show that a typical, low-complexity pure state drawn from any ensemble with large entropy and well-defined effective temperature becomes locally indistinguishable from a Gibbs state upon unitary evolution. In this setting, our rigorous results prove the widely anticipated notion that statistical physics should be understood as an emergent phenomenon, explicitly derived from the first principles of quantum mechanics.
\end{abstract}
\preprint{MIT-CTP/5758}
\maketitle
Establishing a rigorous mathematical foundation for quantum thermalization~\cite{Polkovnikov2011,Gogolin2016,Nandkishore2015,Abanin2019,Alhambra2023} is arguably the most important challenge in quantum statistical physics. Such a feat would position the foundational principle of statistical mechanics --- that a system evolves into a state with maximum entropy --- as a derived, provable statement.
The significance extends beyond this fundamental theory, as quantum thermalization underpins key questions across many other branches of physics, including the equilibration of heavy-ion experiments described by quantum chromodynamics~\cite{Raju2021} and the formation of black holes in certain quantum-gravity theories~\cite{Harlow2016}.  It also relates to practical applications, such as the development of passive quantum memories for quantum computation~\cite{Brown2016}.

The emergence of statistical mechanics in quantum systems is a century-old problem~\cite{vonNeumann2010} which has been attacked by a multitude of complementary approaches, including devising conjectures such as the eigenstate thermalization hypothesis (ETH)~\cite{Deutsch1991,Srednicki1994,Dymarsky2018} and performing numerical or quantum simulations~\cite{Jensen1985,Kinoshita2006,Rigol2008,Trotzky2012,Gring2012,Langen2015,Kaufman2016}. While these approaches provide useful insights and relevant conceptual understanding, they do not constitute the rigorous approach desired to tackle foundational questions. Recent developments in quantum information theory present us with a new avenue. By leveraging symmetries and information theory, a number of rigorous and precise statements about quantum dynamics have been formulated~\cite{Tasaki1998,Goldstein2006,Popescu2006, Reimann2008,Linden2009,Cramer2012,Keating2015,Muller2015,Reimann2016,Mori2016,Farrelly2017,Huang2019,Buca2023,Bertoni2025}, bringing us a step closer to a complete proof of quantum thermalization.

Here, we adopt such a rigorous, information-theoretic approach to identify a broad set of physical conditions under which quantum thermalization must provably occur, without appealing to any unproven hypothesis. We prove that quantum thermalization must occur in any qubit system with local interactions under three conditions:  high effective temperature, translation invariance, and no perfect resonances in the energy spectrum. Unlike previous results, we prove this statement without breaking the locality of the Hamiltonian~\cite{Bertoni2025}, imposing any  assumptions on the energy distribution of the initial state~\cite{Muller2015,Farrelly2017},
or relying on empirical postulates such as the ETH or the emergence of typicality in dynamics~\cite{Popescu2006}.

\section*{Quantum thermalization}

\begin{figure}[t]
    \centering
    \includegraphics[width=1\columnwidth]{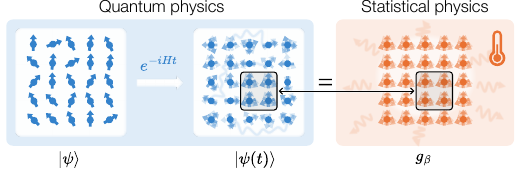}
    \caption{{\bf Main idea of quantum thermalization. }
    The local indistinguishability between time-evolved quantum many-body states (left) and the thermal Gibbs states (right) establishes the bridge between quantum and statistical physics.\vspace{-1em}}
    \label{fig:term}
\end{figure}

We begin by formulating the precise statement of quantum thermalization. In this work, we consider a large but limited class of quantum systems: $N$ qubits arranged on a periodic lattice in arbitrary spatial dimension $D$, with dynamics described by a local, translation-invariant time-independent Hamiltonian $H$. Once the system is initialized in a pure state $\ket{\psi}$ that is far from equilibrium, quantum thermalization dictates that upon  unitary time evolution, the state
\begin{equation}
    \ket{\psi(t)}=e^{-i H t/\hbar}\ket{\psi}
\end{equation} 
becomes locally indistinguishable  from a Gibbs state 
\vspace{-0.5em}
\begin{equation}
    g_\beta = \frac{e^{-\beta H}}{\tr(e^{-\beta H})},
\end{equation}
at a corresponding inverse temperature $\beta = 1/(k_B T)$ determined by the conservation of global energy~$\expval{H}{\psi}=\tr(g_\beta H)$ (in the presence of further conservation laws, $g_\beta$ may be replaced by a generalized Gibbs state). 
This process is depicted in Fig.~\ref{fig:term}. Hereafter, we work with units where the Boltzmann constant $k_B$ and the Planck constant~$\hbar$ equal~$1$. 

Local indistinguishability means equality of local measurement outcomes. Specifically, for any geometrically local and finite region $A$ and for any typical late time $t$, we demand that the evolved state reproduces the thermal expectation values of any observable $O_A$ supported in $A$, 
\begin{equation}
\label{eq:flucs}
   \expval{O_A}{\psi(t)}= \tr(g_\beta O_A) \pm \varepsilon(N),
\end{equation}
where $\varepsilon(N)$ accounts for finite-size corrections~\cite{Srednicki1996,Huang2024}, which must vanish in the thermodynamic limit $N\rightarrow \infty$. Equivalently, we write
\begin{equation}
\label{eq:localtracenormthrm}
\norm{g_\beta-\psi(t)}_A\leq \varepsilon(N),\end{equation} where we introduced the {\it local trace norm} $\norm{\, \cdot \,}_A$, which measures the  distinguishability between $g_\beta$ and    $\psi(t)=\dyad{\psi(t)}$ to an optimal observable in region $A$, i.e.,
$\norm{g_\beta - \psi(t)}_A\coloneqq \max_{O_A} \tfrac{1}{\norm{O_A}_\textrm{op}}\abs{\tr\!\big(\big(g_\beta-\psi(t)\big) O_A\big)}$. We say the state $\psi$ thermalizes under $H$ if Eq.~\eqref{eq:localtracenormthrm} holds. 

To establish a rigorous and universal statement about quantum thermalization, one needs to take into account two important considerations. First, thermalization does not occur in all systems; many examples of non-thermalizing dynamics have been recognized over the past decades~\cite{Basko2006,Pai2019,Sala2020,Khemani2020,Serbyn2021},
and the precise conditions that ensure thermalization or the lack thereof have not been completely identified.
It turns out, as we show here, that only one extra condition on top of translation invariance  suffices to guarantee thermalization of typical states: the energy spectrum $\{E_0,E_1, E_2, \dots\}$  has nondegenerate gaps, meaning that  $E_i+E_j=E_k+E_l$ is exactly satisfied  only if $(i,j)=(k,l)$ or $(i,j)=(l,k)$.
This condition is conjectured to hold for all but a fine-tuned (measure-zero) subset of Hamiltonians, as we further discuss later. 
Violations of this condition imply that one can harvest (potentially hidden) pairs of noninteracting qubit degrees of freedom in the system (see Supplementary Material), which is not expected for ergodic systems. To the best of our knowledge, variants of nondegenerate gap condition have been relied on by all prior  works which rigorously prove quantum equilibration~\cite{Reimann2008,Linden2009,Muller2015,Huang2019}.

The second consideration is that it is provably impossible to devise a  procedure (either formulated as a general theorem or computer algorithm) that determines whether a specific state thermalizes under a given translation-invariant Hamiltonian --- such a task has been shown to be computationally undecidable~\cite{Shiraishi2021}. Therefore, we focus on establishing a   probabilistic statement. We say an ensemble $\mathcal{E}$ consisting of pure states thermalizes if  
\begin{equation}
\label{eq:ensembletherm}
    \E_{\psi}\,\E_t \left[\norm{g_\beta-\psi(t)}_A \right] 
  \leq \varepsilon(N),
\end{equation}
where the average $\E_{\psi}$ is taken over initial states  $\psi$ sampled from the ensemble $\mathcal{E}$ and $\E_t[\,\cdot\,]\coloneq \lim_{T\to \infty} \frac{1}{T}\int_{0}^T \dd{t} [\,\cdot\,]$ denotes averaging over evolution times $t\geq 0$. The local trace norm being averaged  is nonnegative, and therefore Eq.~\eqref{eq:ensembletherm} implies quantum thermalization of all typical initial states $\psi \in\mathcal{E}$ and evolution times $t$.

An important question is to identify for which classes of initial-state ensembles one can prove a meaningful statement about thermalization. 
Generally, we wish to choose the ensemble as large as possible for a broad applicability, yet we note that for certain ensembles the statement of thermalization is less meaningful. For example, if $\mathcal{E}$  contains all states in the Hilbert space uniformly distributed, then thermalization follows trivially, simply because most states in the Hilbert space are already thermal~\cite{Popescu2006}. 
Moreover, such random pure states are not physically realizable either in the laboratory or nature due to their exponentially large complexity~\cite{Poulin2011}.
Instead, the interesting scenario is to consider out-of-equilibrium states that can be easily prepared.
In this direction, a recent work has shown the thermalization of the ensemble of random product states in 1D~\cite{Huang2019}.
While an important milestone, this result only concerns the infinite effective temperature regime $\beta=0$, wherein the ``thermal'' state is independent of the choice of Hamiltonian and energy conservation does not play an important role. 

In order to draw a physically meaningful statement about quantum thermalization at finite temperature, we introduce a class of ensembles $\mathcal{E}_{\beta,\mathcal{C}}$ with two parameters $\beta$ and $\mathcal{C}$ satisfying the following three properties:
\begin{enumerate}
    \item[(P1)] {\it Shallow complexity} $\mathcal{C}$: any state in $\mathcal{E}_{\beta,\mathcal{C}}$ can be prepared from a product state by a quantum circuit of depth at most $\mathcal{C}$, where $\mathcal{C}(N)$ may grow at most subpolynomially in the system size $N$.

    \item[(P2)] {\it Inverse temperature $\beta$}: the average global energy matches the thermal value $\E_{\psi} [\tr(\psi H)] = \tr(g_\beta H)$.

    \item[(P3)]{\it Maximally entropic ensemble} (MEE):  the von Neumann entropy of the average state $\rho=\E_\psi[\psi]$ is maximal among all other ensembles with the same complexity (P1) and  temperature (P2).
\end{enumerate}
 
Property (P1) guarantees that any state in the ensemble is far from equilibrium and physically relevant, (P2) controls the effective temperature of  the ensemble, and (P3) implies that the ensemble is large, in a precise sense. Our main result establishes the quantum thermalization of any ensemble satisfying these three properties at high temperature: 
\begin{theorem}[Quantum thermalization]
\label{th:main}
Consider any translation invariant, geometrically local Hamiltonian for $N$ qubits in $D$-dimensional lattice with nondegenerate spectral gaps. Below a certain threshold inverse temperature $\beta_{*}>0$ (independent of $N$), any ensemble satisfying properties (P1--3)
 thermalizes \textup{[}i.e., satisfies Eq.~\eqref{eq:ensembletherm}\textup{]}, with  finite-size corrections $\varepsilon(N)\leq O\big(N^{-{1/2}+\delta}\big)$ for any $\delta>0$ and geometrically local region $A$ of constant size. 
\end{theorem}

In particular, if we consider an ensemble of trivial complexity $\mathcal{C}=0$, we establish a statement about thermalization of initial product states:
\begin{corollary}
Under the conditions of Theorem~1, any maximally entropic ensemble of pure product states at inverse temperature $\beta\leq \beta_*$ thermalizes.
\end{corollary}

\begin{figure}
    \centering
    \includegraphics[width=\columnwidth]{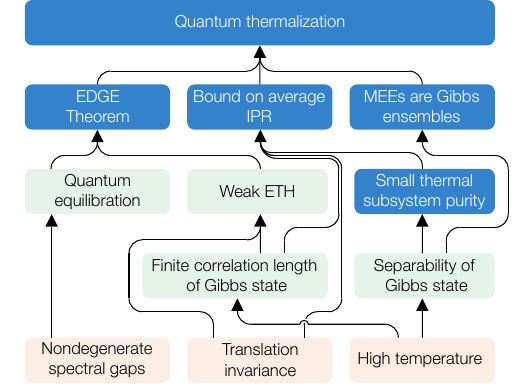}
    \caption{{\bf Flowchart for our proof of quantum thermalization}. We summarize logical implications among our results (blue), previously-known results (green), and assumptions (orange).}
    \label{fig:log-flow}
\end{figure}

The proof of Theorem 1 is enabled by synthesizing a number of previously-established and newly-developed technical results, summarized by the flowchart in Fig.~\ref{fig:log-flow}. Among them, three particular results are pivotal. First, 
we introduce a notion that we call 
\emph{energy dispersed Gibbs ensembles} (EDGEs) --- these are ensembles of pure states, wherein individual states are typically supported in many energy eigenstates (energy dispersed), while their collection averages to the Gibbs state (Gibbs ensemble).
By adapting previous results (the weak ETH~\cite{Biroli2010,Mori2016,Brandao2019,Alhambra2020} and quantum equilibration theorems~\cite{Linden2009}), our EDGE Theorem shows that EDGEs thermalize under the nondegenerate spectral gaps condition, as long as the Gibbs state has finite correlation length, which provably occurs at high enough temperature~\cite{Kliesch2014} $|\beta| < \beta_*^{\textrm{(corr)}}$. Second, we note that any maximally entropic ensemble of shallow-complexity states at high temperature is a Gibbs ensemble. This follows from a recent result regarding the separability of Gibbs states~\cite{Bakshi2024}, which proves the existence of a Gibbs ensemble consisting of product states at sufficiently high temperature $ |\beta| < \beta_*^{(\mathrm{sep})}$. Finally, using the translation invariance, we show that any high-temperature Gibbs ensemble of shallow-complexity states must also be energy dispersed, and thus constitute an EDGE. We now give an overview of these  results, with full technical details available in the Supplementary Material. 

\section*{EDGE Theorem}
We begin by formally introducing EDGEs.
\begin{definition}[Energy dispersed Gibbs ensemble] For any $N$-qubit Hamiltonian $H$ with Gibbs state $g_\beta$, an ensemble of states $\mathcal{E}_\beta$ is an EDGE if it satisfies the following two conditions:
\label{def:EDGE}
\begin{enumerate}
    \item[(a)] [Gibbs ensemble] The average of the ensemble is close to $g_\beta$,
    \begin{equation}
        \Big\|\E_{\psi}[\psi]- g_\beta\Big \|\leq2^{-N}\varepsilon_{\mathrm{GE}}(N),
    \end{equation} where $\norm{\,\cdot\,}$ is the usual trace norm.
    \item[(b)][Energy dispersion] The average inverse participation ratio (IPR) in the energy eigenbasis $\{\ket{E_j}\}_j$ is upper bounded as
    \begin{equation}
         \E_{\psi}\Big[\sum_{j=1}^{2^N} \abs{\braket{E_j}{\psi}}^4\Big]\leq\varepsilon_{\mathrm{ED}}(N).
    \end{equation}
\end{enumerate}
  The two functions $\varepsilon_{\mathrm{ED}}(N)$ and $\varepsilon_{\mathrm{GE}}(N)$ should be nonnegative and vanish in the thermodynamic limit $N\to \infty$, but otherwise can be arbitrary.  
\end{definition}

 The dispersion in the energy eigenbasis is measured by a small IPR $\sum_j\abs{\braket{E_j}{\psi}}^4$, which is the inverse of the so-called effective dimension or participation ratio, that estimates the number of eigenstates on which $\ket{\psi}$ is supported.

Our next result shows that typical states in an EDGE thermalize as long as the Gibbs state has finite correlation length. Finite correlation length means that correlations between any pair of observables decay at least exponentially fast in their distance (up to a boundary-dependent factor). This condition holds below a certain constant inverse temperature $\beta_*^{\mathrm{(corr)}}$ for any local Hamiltonian in any geometry~\cite{Kliesch2014} (see Supplementary Material).

\begin{theorem}[EDGE Theorem]
\label{th:EDGE}
Consider any translation invariant, geometrically local Hamiltonian for $N$ qubits in $D$-dimensional lattice with nondegenerate spectral gaps.
    Let $\mathcal{E}_\beta$ be an EDGE and assume that the Gibbs state $g_\beta$ has finite correlation length.
    Then, for any geometrically local  region $A$ with a constant number of qubits $N_A$,  \begin{align}
    \nonumber
        \E_{\psi}& \E_t \left[\norm{g_\beta-\psi(t)}_A \right] \leq\\& O(N^{-\gamma})+2\varepsilon_{\mathrm{GE}}(N)+{2^{N_A}} \varepsilon_{\mathrm{ED}}(N)^{1/2}, \label{eq:boundav}
    \end{align}
for any $\gamma<1/(D+1)$ at any temperature and for any $\gamma<1/2$ at sufficiently high temperature.
\end{theorem}
 Our proof of Theorem~\ref{th:EDGE} is inspired by Ref.~~\cite{Huang2019}. Below we present the key ideas, with the full  technical details available in the  Supplementary Material.  

\begin{proof} Quantum thermalization, $\norm{\psi(t)-g_\beta}_A \to 0$, can be split into two fine-grained statements.
First, we show equilibration, which dictates that pure quantum states at late times become locally indistinguishable from their time-averaged mixed state $\rho_\infty\coloneqq \E_\tau[\psi(\tau)]$, i.e., $\norm{\psi(t)-\rho_\infty}_A \to 0$.
Second, we prove a property akin to ergodicity, the local equivalence between temporally and thermally averaged ensembles $\norm{\rho_\infty-g_\beta}_A \to 0$. Once these two statements are proven, one obtains the desired result via a triangle inequality.

A sufficient condition for quantum equilibration is well known when assuming nondegenerate spectral gaps: the vanishing of the IPR~\cite{Linden2009}, leading to 
\begin{equation}
    \E_{\psi}\,\E_t \left[\norm{\psi(t)-\rho_\infty}_A \right] \leq {2^{N_A}} \E_{\psi}\Big[\sum_{j=1}^{2^N} \abs{\braket{E_j}{\psi}}^4\Big] ^{1/2}, 
    \nonumber \label{eq:IPRBound}
\end{equation} 
which gives the last term in Eq.~\eqref{eq:boundav} upon using property~(b) of an EDGE. 

To prove the equivalence of ensembles, we utilize a modified version of a seminal result, called the weak eigenstate thermalization hypothesis (weak ETH)~\cite{Biroli2010,Mori2016,Brandao2019,Alhambra2020}. 
\begin{prop}[A corollary of the weak ETH]
\label{prop:weaketh}
     Consider any translation invariant, geometrically local Hamiltonian for $N$ qubits in a $D$\nobreakdash-dimensional periodic lattice, with eigenbasis $\{\ket{E_j}\}_j$ which is common to the translation operator. Assume that the Gibbs state $g_\beta$ has finite correlation length and consider a geometrically local region $A$ with a constant number of qubits.  Then, \begin{equation}
        \sum_{j=1}^{2^N} \expval{g_\beta}{E_j} \big\|\dyad{E_j} - g_\beta\,\big \|_A\leq O(N^{-\gamma}),
    \end{equation} 
    for any $\gamma<1/(D+1)$ at any temperature and for any $\gamma<1/2$ at sufficiently high temperature.
\end{prop}
The proof of Proposition~\ref{prop:weaketh}, which leverages results from Refs.~\cite{Anshu2016,Alhambra2020,Kuwahara2020}, is available in the  Supplementary Material.
By applying this result to $\rho=g_\beta$, we establish the equivalence between the thermal and temporal ensembles with bounded error:
\begin{align}
\E_\psi \norm{\rho_\infty-g_\beta}_A &= 
\E_\psi \Big\| \sum_j \abs{\braket{E_j}{\psi}}^2 (\dyad{E_j} - g_\beta)  \Big\|_A
\nonumber
\\
&\leq \sum_{j}\E_{\psi} \big[ \expval{\psi}{E_j}\big] \big\|\dyad{E_j} - g_\beta\big\|_A  \nonumber \\
&\leq \sum_j p_j \big\| \dyad{E_j} - g_\beta\big\|_A + 2\varepsilon_{\mathrm{GE}}(N),
\nonumber
\end{align}
where we used $\rho_\infty = \sum_j  \abs{\braket{E_j}{\psi}}^2  \dyad{E_j} $ for a given $\ket{\psi}$ (owing to the nondegenerate spectral gap condition) in the first line, the triangle inequality in the second line, and the substitution of $\E_\psi[\psi] \approx g_\beta$ up to a bounded error $2^{-N} \varepsilon_{\mathrm{GE}}(N)$ 
in the third line.
This completes the proof.
\end{proof}
In the Supplementary Material we present generalizations of the EDGE theorem to systems with additional conserved quantities, generalized Gibbs ensembles~\cite{Vidmar2016} and finite time~\cite{Short2012}.

\section*{EDGEs of shallow complexity at finite temperature}
While we have proven the thermalization of typical states in EDGEs, it remains to show explicit examples or at least the existence of these ensembles under physical conditions. 
At infinite temperature, they can be easily constructed ---
the random product-state ensemble
constitutes an EDGE, regardless of the Hamiltonian. 
At finite temperature, however, such ensembles necessarily depend on the details of the Hamiltonian. In what follows, we prove that EDGEs of shallow complexity always exist at high temperature.
We establish this in two steps.
First, we show that, at high enough temperature, any maximally entropic ensemble (P3) of any complexity $\mathcal{C}$ (including the most stringent case $\mathcal{C}=0$) are Gibbs ensembles.
Second, we show that whenever $\mathcal{C}(N)$ is shallow (P1), such ensemble is energy dispersed.
\begin{lemma}
\label{lemm:HT-MEE-GE}
For any geometrically local Hamiltonian and any choice of complexity $\mathcal{C}(N)$, there exists a constant threshold  $\beta_*^{(\mathcal{C})}\geq \beta_*^{(\mathrm{sep})}$ such that any maximally entropic ensemble  $\mathcal{E}_{\beta,\mathcal{C}}$ with complexity $\mathcal{C}$ and inverse temperature $|\beta|\leq \beta_*^{(\mathcal{C})}$ averages to the Gibbs state, $\E_{\psi}[\psi]=g_\beta$.
Furthermore, the threshold temperature does not increase as the complexity increases: $\beta_*^{(\mathcal{C})} \geq \beta_*^{(\mathcal{C'})}$ if $\mathcal{C}\geq \mathcal{C'}$.
\end{lemma}

The statement follows directly from a recent result from quantum information theory. Reference~~\cite{Bakshi2024} proves that the Gibbs state of any local Hamiltonian below a threshold inverse temperature $\beta_*^{(\mathrm{sep})}>0$ is separable, by an explicit construction of a product-state Gibbs ensemble. Because the Gibbs state uniquely maximizes the von Neumann entropy, this ensemble is a maximally entropic ensemble with complexity $\mathcal{C}=0$ and thus any other maximally entropic ensemble at the same temperature also must average to same unique Gibbs state.
As the complexity constraint is relaxed away from product states, the threshold temperature cannot increase. It is not clear if it strictly decreases, i.e., if the Gibbs state at lower temperatures can be unraveled into an ensemble of low complexity states. This question deserves future investigation. 
 
Now we turn to show the energy dispersion, for which we leverage the translation invariance.
\begin{theorem}[Bound on average IPR]
\label{th:boundonIPR}
     In a periodic lattice of $N$ qubits, consider an ensemble $\mathcal{E}$ of pure states with shallow complexity $\mathcal{C}$, whose average state $\rho=\E_{\psi\in \mathcal{E}}[\psi]$ satisfies (i)~translation invariance, (ii)~finite correlation length, and (iii)~exponentially small subsystem purity $[\tr(\rho_A^2)\leq e^{-\Omega(N_A)}]$. Then, the average IPR in an eigenbasis $\{\ket{j}\}_j$ of the translation operator is bounded,
    \begin{equation}
    \label{eq:boundofIPR}
        \E_{\psi}\Big[\sum_{j=1}^{2^N} \abs{\braket{j}{\psi}}^4\Big]\leq O(N^{-1+\delta})
    \end{equation} for any $\delta>0$.
\end{theorem}

We prove in the Supplementary Material that the three properties required on $\rho$ in Theorem~\ref{th:boundonIPR} are satisfied by the Gibbs state at high temperatures $|\beta|<\beta_*$, so taking $\ket{j}=\ket{E_j}$, and combining
Theorem~\ref{th:boundonIPR},
Lemma~\ref{lemm:HT-MEE-GE} and the EDGE Theorem we obtain our main result, Theorem~\ref{th:main}.  While the detailed proof of Theorem~\ref{th:boundonIPR} is presented in the Supplementary Material, below we illustrate the key intuition by proving a simplified version in 1D.
\begin{prop} \label{prop:compbasis}The average IPR on any eigenbasis $\ket{j}$ of the translation operator over the uniform ensemble of computational basis states $\ket{b}$ with $b \in \{0,1\}^{N}$ is upper bounded
    \begin{align}
\label{eq:boundIPRcomstates}
         \E_{b\in \{0,1\}^{N}}\Big[\sum_{j=1}^{2^N}\abs{\braket{j}{b}}^4\Big]\leq \frac{1}{N} + O(2^{-N/2}).
     \end{align}
\end{prop}

Note that Proposition~\ref{prop:compbasis} combined with the EDGE Theorem already proves a novel result: that a random computational basis state will typically thermalize to infinite temperature for any translation-invariant Hamiltonian with nondegenerate spectral gaps in 1D.

\begin{figure}
    \centering
    \includegraphics[width=1\columnwidth]{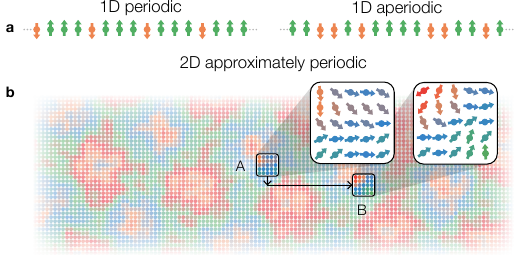}
    \caption{{\bf Schematic illustrations of product states on 1D or 2D lattices.}
    {\bf a} Periodic (left) and aperiodic (right) computational basis states in 1D.
    {\bf b}~Approximately periodic product state in 2D: qubit configurations in regions A and B separated by a period are similar, giving rise to long-range  correlations. } 
    \label{fig:periodicity}
\end{figure}

\begin{proof}
          The central idea is to divide the bitstrings ${b\in\{0,1\}^{N}}$ that enumerate the states into two different groups and bound their contributions to the average IPR separately: periodic bitstrings, which are invariant under some nontrivial translation on the circle, and aperiodic ones, which are not equal to any of their nontrivial translations  (see Fig.~\ref{fig:periodicity}a).
          The former group constitutes a vanishingly small fraction among all $2^N$ possibilities, hence their contribution becomes small. Specifically, since the period cannot be longer than $N/2$, there are at most $O(2^{N/2})$ distinct periodic bitstrings, leading to the second term in Eq.~\eqref{eq:boundIPRcomstates}.
          The second, aperiodic group contains the majority of the bitstrings, but their overlaps with the translation operator eigenstates are necessarily small.
          Specifically, for $N$ distinct bitstrings $b_1, b_2, \dots, b_N$ which are translations of each other, their overlaps with an translation operator eigenstate must be the same up to a global phase, resulting in a bound $\abs{\braket{j}{b}}^2\leq 1/N$. This leads to the first term in Eq.~\eqref{eq:boundIPRcomstates}, completing the proof.
\end{proof}

 Theorem~\ref{th:boundonIPR} is proven by extending the previous argument to arbitrary shallow-complexity states defined on 1D or higher-dimensional lattices.
 This generalization involves solving two nontrivial complications.
 First, unlike the bitstring case, one cannot sharply differentiate periodic versus aperiodic states because generic states have a nonzero overlap with their translations. This is overcome by introducing {\it approximately periodic states}, which are those states whose overlap with some translation is above a certain threshold (Fig.~\ref{fig:periodicity}b). 
 Second, in general bounding the contribution to the IPR from approximately periodic states requires going beyond simple counting. For this we utilize property (ii), the finite correlation length of $\rho$. Intuitively, approximately periodic states cannot be weighted by high probabilities in the ensemble because otherwise they would induce long-range correlations in $\rho$ (see Fig.~\ref{fig:periodicity}). This argument implicitly requires  that the local regions $A$ and $B$ have a sufficient amount of entropy to support the correlation, hence property (iii).

\section*{Generalized Gibbs state and finite-time thermalization}
So far we have focused on thermalization to the canonical ensemble, where the Gibbs state is specified by only one conserved quantity: energy. In the Supplementary Material we present  generalizations of our results for thermalization to generalized Gibbs states $g_{\{\lambda_i\}}\propto \exp(-\sum_i \lambda_i Q_i)$ with a constant number of conserved quantities $[Q_i,H]=0$. An analogous statement to Theorem 1 holds if $\max_i|\lambda_i|\leq \lambda_*$ for a certain threshold $\lambda_*$ which is independent of the system size (a generalization of {\it high temperature}). Our result is applicable also to certain integrable systems~\cite{LevkovichMaslyuk2016}, as long as the initial state is drawn from a generalized maximally entropic ensemble in which only a finite number of generalized chemical potentials $\lambda_i$ are nonzero.

Furthermore, while so far we have only discussed thermalization in the infinite-time limit, our work can be improved to provide finite-time statements leveraging the results from Ref.~\cite{Short2012}. In the Supplementary Material we present a bound on the thermalization time in terms of the inverse of the minimal {\it gap of gaps} which is expected to be exponentially small in the system size.
Such exponential dependence might me unavoidable in our setting, as recent developments identify translation-invariant systems which can thermalize only after exponentially long timescales~\cite{Balasubramanian2024}.
 It is an important question to identify what additional physical  conditions can guarantee a fast quantum thermalization in closed quantum systems.

\section*{Discussion}
It has been widely expected that statistical mechanics should be understood as an emergent phenomenon out of a microscopic quantum description~\cite{Gogolin2016,Wallace2021,Popescu2006}. Our rigorous results prove this expectation in a  large class of models.
We stress that even at high temperature where the Gibbs state is separable, quantum thermalization necessitates a large amount of entanglement. The system is initialized in a low-entanglement state which eventually becomes locally indistinguishable from the volume-law entropic Gibbs state, implying that the entanglement entropy of any subsystem grows over time and saturates to a volume-law scaling.

Our Theorem~\ref{th:main} establishes a  universal statement of quantum thermalization which is agnostic of the details of the Hamiltonian. In practice, however, one often has access to more fine-grained information, and our several other  technical results might be useful. For example, in 1D systems, the Gibbs state always has a finite correlation length, so the the EDGE theorem is applicable at any temperature.

Our work leaves several other questions and generalizations open. First, in going beyond translation invariant systems, one open question is whether one can prove a version of the weak ETH in the presence of weak disorder. However, strongly disordered systems in 1D systems are largely discussed to exhibit many-body localization~\cite{Basko2006,Nandkishore2015,Abanin2019}, which, if exists, would constitute a robust violation of quantum thermalization.

Second, it would be desirable to  prove that the nondegenerate gap condition is generically satisfied. Unlike the translation invariance and the high temperature conditions, the nondegenerate gap condition is hidden in the energy spectrum, and hence difficult to check. Nevertheless, it is expected to  generically hold true, in a precise sense.
Violations of the nondegenerate spectral gap condition are proven to be zero-measure for arbitrary local Hamiltonians~\cite{Huang2021}, implying that those are rare, fine-tuned exceptions.
For translation-invariant Hamiltonians, the same is conjectured to hold for every system size $N$~\cite{Huang2019}, but has not yet been rigorously proven. A breakdown of this conjuncture would have counterintuitive consequences: that for some values of $N$,  every translation-invariant Hamiltonian  necessarily has at least one gap degeneracy and consequently hidden noninteracting degrees of freedom. This seems unlikely, and has been already ruled out by explicit numerical tests for relatively small $N$~\cite{Muller2015}.

Third, it is important to prove thermalization at lower temperatures.
Away from critical points, even at low temperature, the thermal system may be described by (a mixture of a few) Gibbs state(s) with finite correlation length. A nontrivial task is to show whether or not such Gibbs states can be unraveled into an ensemble of pure shallow-complexity states.
This structural question is very interesting on its own.

More broadly, one should investigate the thermalization of more general physical systems beyond qubits on lattices, such as bosonic particles with unbounded local Hilbert-spaces, those possessing non-abelian symmetries (e.g., arising from spatial inversion), and quantum field theories; with the ultimate aim of 
constructing a universal framework explaining the emergence of statistical mechanics in quantum systems.

\begin{acknowledgments}
We thank 
A.~M.~Alhambra,
A.~Anshu,
A.~Bakshi,
S.~Balasubramanian,
F.~Brand\~ao,
J.~Eisert,
J.~Haferkamp,
A.~Harrow,
W.~W.~Ho,
Y.~Huang,
I.~H.~Kim,
A.~Liu,
Y.-J.~Liu,
D.~Mark, 
D. Ranard,
and 
P. Reimann
for insightful conversations and feedback on earlier versions of this preprint. In particular, we acknowledge A.~M.~Alhambra for a detailed discussion at the initial stage of this work and Y.~Huang for a careful explanation of the techniques used in Ref.~~\cite{Huang2019}, some of which are also utilized in the present work. This work was supported by the Center for Ultracold Atoms (an
NSF Physics Frontiers Center; PHY-2317134), NSF CAREER (DMR-2237244), and the Heising-Simons Foundation (grant \#2024-4851). S.C. also acknowledges support from the Sloan Foundation through
a Sloan Research Fellowship.
\end{acknowledgments}

\bibliography{references}
\bibliographystyle{naturemag}

\end{document}


\def\bibsection{\section*{\refname}} 
\newcommand{\E}[0]{\mathop{{}\mathbb{E}}}
\newcommand{\Pro}[0]{\mathop{{}\mathbb{P}}}
\title{Supplementary Material: 
Quantum thermalization must occur \\in translation-invariant systems at high temperature}
\author{Sa\'ul Pilatowsky-Cameo}
\email{saulpila@mit.edu}
\affiliation{Center for Theoretical Physics, Massachusetts Institute of Technology, Cambridge, MA 02139, USA}

\author{Soonwon Choi}
\email{soonwon@mit.edu}
\affiliation{Center for Theoretical Physics, Massachusetts Institute of Technology, Cambridge, MA 02139, USA}
 \newtheorem{theorem}{Theorem}

    \newtheorem{corollary}{Corollary}
    \newtheorem{lemma}{Lemma}
    \newtheorem{prop}{Proposition}
    \newtheorem{conjecture}{Conjecture}
\theoremstyle{definition}
  \newtheorem{definition}{Definition}
\def\theequation{S\arabic{equation}}
\renewcommand{\thecorollary}{S\arabic{corollary}}
\renewcommand{\thelemma}{S\arabic{lemma}}
\renewcommand{\theprop}{S\arabic{prop}}
\renewcommand{\thedefinition}{S\arabic{definition}}
\renewcommand{\theconjecture}{S\arabic{conjecture}}
\renewcommand{\thetheorem}{S\arabic{theorem}}
\newcommand{\maintextref}[1]{{#1}} 
\renewcommand{\figurename}{Supplementary Figure}

\maketitle

In this Supplementary Material, we provide detailed information, complete rigorous proofs, and generalizations of our results.
In Section~\ref{sec:preliminaries}, we introduce our notations, the physical setup under consideration, a detailed discussion of the nondegenerate spectral gap condition, and some useful properties of Gibbs states and shallow-complexity states to be used in later sections.
Sections~\ref{sec:EDGE}, ~\ref{sec:MEE}, and ~\ref{sec:boundonaverageIPR} are dedicated to formally present and prove our technical theorems outlined in the main text.
Specifically, Section~\ref{sec:EDGE} contains the detailed proof of our EDGE Theorem,  Section~\ref{sec:MEE} presents the detailed definition of maximally entropic ensembles and their connection to Gibbs ensembles at high temperature, and Section~\ref{sec:boundonaverageIPR} contains the detailed proof of our bound on the average IPR, Theorem 3. Finally, in Section~\ref{sec:generalized} we generalize our results to generalized Gibbs states in the presence of additional conserved quantities.

\tableofcontents

\section{Preliminaries}
\label{sec:preliminaries}
\subsection{Setting and notations}
We begin by introducing our physical setup and general notations.
\subsubsection{Asymptotic notation}
We make use of the big-$O$ and related asymptotic notations. When we write $f(N)\leq O(g(N))$, we mean that there exist constants $C>0$ and $N_*>0$ such that $f(N)\leq C g(N)$ for all $N\geq N_*$. We sometimes also write $f(N)\leq e^{O(g(N))}$, which means $f(N)\leq e^{C g(N)}$ for all $N\geq N_*$. For lower bounds, we write $f(N)\geq \Omega(g(N))$ which means $f(N)\geq  Cg(N)$ for all $N\geq N_*$ and  $f(N)\leq e^{-\Omega(g(N))}$ means $f(N)\leq e^{-C g(N)}$ for all $N\geq N_*$. Finally, $f(N)=\Theta(g(N))$  means  that both the upper bound $f(N)\leq O(g(N))$ and lower bound $f(N)\geq \Omega(g(N))$ are satisfied.

We remark that these relations are typically expressed using equality rather than inequality signs. However, we find that the inequalities provide additional clarity.

\subsubsection{Lattice}
We consider systems defined over a $D$-dimensional lattice $\Lambda=\{0,1,2,\dots,L-1\}^D$,  consisting of $N=L^D$ sites.
As it will become clear, the specific geometric properties such as the aspect ratios of the lattice are not important for any of our results.
We will specialize to periodic boundary conditions, so we will think of $\Lambda$ as a torus. Given a point $\bm x\in\Lambda$, with coordinates $\bm x=(x_1,\dots,x_D)$, we measure its magnitude via the Manhattan norm,
\begin{align}
    \norm{\bm x}_1=\sum_{j=1}^D\min\{x_j,L-x_j\}.
\end{align}
A pair of positions in the lattice can be added or subtracted in the conventional way, i.e. with the entrywise modular addition and subtraction.
The distance between two sites $\bm x, \bm y\in \Lambda$ is given by $\norm{\bm x-\bm y}_1$. 

We will study small regions (subsets) of the lattice, in the following sense.
\begin{definition}[Geometrically local region]
    We say a region $A\subseteq \Lambda$ is  \emph{ geometrically local} if its diameter is bounded  by a constant \begin{equation}
        \max_{\bm x, \bm y\in A}\norm{\bm x-\bm y}_1\leq O(1).
    \end{equation} 
\end{definition}

\subsubsection{Hilbert space, observables and Hamiltonian}
On each lattice site, we place a single qubit, i.e. a spin-$1/2$ particle, with local Hilbert space $\mathcal{H}_{\bm x}=\mathbb{C}^2$. The global Hilbert space is $\mathcal{H}=\bigotimes_{\bm x\in \Lambda} \mathcal{H}_{\bm x}$. We say an operator $O$ on $\mathcal{H}$ acts on a region $A\subseteq \Lambda$ (or is supported on $A$) if there is an operator $O_A$ acting on $\mathcal{H}=\bigotimes_{\bm a\in A} \mathcal{H}_{\bm a}$ such that $O=O_A\otimes \bigotimes_{\bm x\not \in A} \mathds{1}_{\bm x}$. Henceforth, we abuse notation and not distinguish between $O_A$ and~$O$ whenever the latter is supported in $A$.

For a $\bm z\in \Lambda$, the operator $\mathbb{T}^{\bm z}$ is the unitary translation operator, which maps a product state $\ket{\psi}=\bigotimes_{\bm x\in \Lambda} \ket{\psi_{\bm x}}$ to 
$\mathbb{T}^{\bm z}\ket{\psi}=\bigotimes_{\bm x\in \Lambda} \ket{\psi_{\bm x-\bm z \mod L}}$. We study geometrically $k$-local, translation-invariant (TI) Hamiltonians $H$, which means that $H= \sum_i h_i $, where each $h_i$ acts on a geometrically local region containing at most $k$ qubits and has bounded operator norm, and $H=\mathbb{T}^{\bm z}H\mathbb{T}^{-{\bm z}}$ for any $\bm{ z}\in  \Lambda$. We will often just say $H$ is local, when the number $k$ is not important. 

\subsubsection{Local trace norm}
We measure the local indistinguishability of two states with the following notion.
\begin{definition}[Local trace norm]
    Given an operator $W$, its trace norm $\norm{W}$ is equal to the sum of the singular values of $W$. Given a region $A\subseteq \Lambda$ with complement $\bar{A}$, we denote 
    $\norm{W}_A=\norm{\tr_{\bar{A}}(W)}$.\end{definition}

From the properties of the trace norm, one can see that
\begin{equation}
    \norm{\rho -\sigma}_A=\max_{\norm{O_A}_{\mathrm{op}}\leq 1} \abs{\tr(\rho O_A)-\tr(\sigma O_A)},
\end{equation}
where $O_A$ runs over all observables supported on $A$ with operator norm  $\norm{\,\cdot\,}_{\mathrm{op}}$ bounded by $1$. Technically, $\norm{\,\cdot\,}_A$ is not a norm, but only a seminorm as  $\norm{\rho -\sigma}_A=0$ implies only that the states $\rho$ and $\sigma$ are locally indistinguishable $ \tr_{\bar{A}}(\rho)=\tr_{\bar{A}}(\sigma)$, meaning that they cannot be discriminated using observables supported on region $A$.

\subsection{Nondegenerate spectral gaps }
One of the most important, nontrivial properties that we demand from the spectrum of Hamiltonians is the \emph{nondegenerate spectral gaps} condition.
\begin{definition}[Nondegenerate spectral gaps]
    A Hamiltonian has \emph{nondegenerate spectral gaps} if its spectrum $\{E_j\}_j$ has no degeneracies and no gap degeneracies, meaning that
    \begin{equation}
    \label{eq:nondegenerategapcondition}
        E_i+E_j=E_m+E_l \iff  (i,j)=(m,l) \text{ or } (i,j)=(l,m).
    \end{equation}
\end{definition}
\noindent This condition is also called the second no-resonance condition \cite{Mark2024}. 
\subsubsection{Physical meaning of the nondegenerate spectral gap condition}

The nondegenerate gap condition has physical interpretation: it rules out  the existence of hidden noninteracting degrees of freedom, as follows.
\begin{prop}[Gap degeneracies arise because of hidden noninteracting qubits]
\label{prop:gapdegeneraciescomfromnoninteractingdegreesoffreedom}
If a Hamiltonian $H$ violates the nondegenerate gap condition, then there exists a hidden pair of noninteracting qubits, in the following sense. One can find a four-dimensional  subspace $\widetilde{\mathcal{H}}\cong \mathbb{C}^2\otimes \mathbb{C}^2\subseteq \mathcal{H}$ and its orthogonal complement $\widetilde{\mathcal{H}}_{\perp}$ such that the Hamiltonian decomposes into  $H=\widetilde{H}+\widetilde{H}_\perp$, where   (i) $\widetilde{H}$ acts only within $\widetilde{\mathcal{H}}$,
(ii) $\widetilde{H}_\perp$ acts only within $\widetilde{\mathcal{H}}_{\perp}$, and 
(iii) $\widetilde{H}$ describes  two noninteracting qubits:
\begin{equation}
\label{eq:Hamnonintsubspace}
\widetilde{H} =a\,\widetilde{Z}_1+b\,\widetilde{Z}_2 +c.
\end{equation}
Here, $a,b,c$ are constants and $\widetilde{Z}_i$ are Pauli-$Z$ operators acting on one of the two tensor factors  which compose $\widetilde{\mathcal{H}}$.  
\end{prop}
\noindent The significance of Proposition S1  comes from the fact that $\widetilde{H}$ lacks an interaction term such as $ \widetilde{Z}_1\widetilde{Z}_2$. We note that the two qubits in $\widetilde{\mathcal{H}}$ are not necessarily physical qubits, and in general $\widetilde{Z}_1$ and $\widetilde{Z}_2$ might be highly nonlocal operators in the lattice. This phenomenon is conceptually similar  to  the hidden noninteracting degrees of  freedom that arise in free-fermionic spin-chains.
\begin{proof}
    Let $\widetilde{H}$ be a subspace spanned any four different energy eigenstates $\ket{E_i}, \ket{E_j},\ket{E_m},\ket{E_l}$, i.e. with all indices different. We identify these states with the canonical basis of $\mathbb{C}^2\otimes \mathbb{C}^2$, as follows
    \begin{align}
        |\widetilde{00}\rangle&=\ket{E_m}, &|\widetilde{01}\rangle&=\ket{E_i}, &|\widetilde{10}\rangle&=\ket{E_j}, &|\widetilde{11}\rangle&=\ket{E_l}.
    \end{align}
    We add the symbol $\widetilde{\cdots}$ to emphasize that these are not physical spin states. We consider the Pauli operators \begin{align}
        \label{eq:paulimattransformed}\widetilde{Z}_1&=\dyad{\smash{{\widetilde{00}}}}+\dyad{\smash{\widetilde{01}}}-\dyad{\smash{\widetilde{10}}}-\dyad{\smash{\widetilde{11}}} &&\text{and}&
        \widetilde{Z}_2&=\dyad{\smash{{\widetilde{00}}}}-\dyad{\smash{\widetilde{01}}}+\dyad{\smash{\widetilde{10}}}-\dyad{\smash{\widetilde{11}}}.
    \end{align}
    By substituting Eq.~\eqref{eq:paulimattransformed} into Eq.~\eqref{eq:Hamnonintsubspace} and equating the result with   \begin{equation}
H|_{\widetilde{\mathcal{H}}}=E_m\dyad{\smash{\widetilde{00}}}+E_i\dyad{\smash{\widetilde{01}}}+E_j\dyad{\smash{\widetilde{10}}}+E_l\dyad{\smash{\widetilde{11}}},
    \end{equation} 
    we obtain the following set of equations
    \begin{align}
        E_m&=a+b+c,&E_i&=a-b+c,&E_j&=-a+b+c,&E_l&=-a-b+c,
    \end{align}
    which have a solution if and only if $E_i+E_j=E_m+E_l$, in which case $a=\tfrac{1}{2}(E_j-E_m)$, $b=\tfrac{1}{2}(E_j-E_l)$, and $c=\tfrac{1}{2}(E_l+E_m)$. Because we are assuming all indices different, the equation $E_i+E_j=E_m+E_l$ amounts to a violation of the nondegenerate spectral gap condition, which is thus equivalent to  the Hamiltonian $H|_{\widetilde{\mathcal{H}}}$ having the form in Eq.~\eqref{eq:Hamnonintsubspace}.
\end{proof}

\subsubsection{Genericity of nondegenerate spectral gaps}
The nondegenerate spectral gap condition is believed to hold for generic Hamiltonians. It has been shown that all Hamiltonians violating this condition form a zero measure set in the space of $k$-local Hamiltonians on $N$ qubits for all $N$~\cite{Huang2021}.
Even within the space of {\it translation-invariant} Hamiltonians on $N$ qubits, it is conjectured that all exceptions form a zero measure set. The latter space can be parametrized, for example, as
\begin{equation}
\label{eq:all-k-local-hams}
    H=\sum_{\bm n \in \Lambda } \mathbb{T}^{\bm n}\sum_P J_P P\mathbb{T}^{-\bm n},
\end{equation}
where $P$ runs over all Pauli strings which act on some fixed geometrically local  region of $k$ qubits, and   the coefficients are $J_P\in [-1,1]$. 
\begin{conjecture}[Conjecture 2 of Ref.~\cite{Huang2019}: almost all $k$-local TI Hamiltonians have nondegenerate gaps]\label{conj:noresonanceconditionsaregeneric} For a fixed $k\geq 2,N,$ and $D$, almost all (geometrically) local TI Hamiltonians on the $D$-dimensional periodic lattice with $N$ qubits have nondegenerate spectral gaps, with respect to the ensemble produced by taking the coefficients $J_P$ in  Eq.~\eqref{eq:all-k-local-hams} to be uniformly distributed in $[-1,1]$.
\end{conjecture}

Although proving conjecture \ref{conj:noresonanceconditionsaregeneric} remains an open problem, Ref.~\cite{Huang2019} has made some progress by showing the following:
\begin{theorem}[Lemma 8 of \cite{Huang2019}]
    For a fixed $k\geq 2,N,$ and $D$, one (and only one) of the following statements holds:
    \begin{enumerate}
        \item Almost all $k$-local TI Hamiltonians on the $D$-dimensional periodic lattice with $N$ qubits have nondegenerate spectral gaps. 
        \item No $k$-local TI Hamiltonian on the $D$-dimensional periodic lattice with $N$ qubits has nondegenerate spectral gaps.
    \end{enumerate}
\end{theorem}
\noindent 

Conjecture \ref{conj:noresonanceconditionsaregeneric} asserts that case 1 in the above theorem is the true statement. 
Thus, proving the conjecture only requires showing that case 2 does not  occur, which reduces to finding a single example of a $k$-local TI system with nondegenerate spectral gaps. This has been done numerically~\cite{Muller2015} for accessible system sizes.
For all systems sizes $N$ checked so far, case 1 holds as expected.
However, for arbitrary system sizes, it appears difficult to attain analytical understanding of the spectrum of a Hamiltonian  with nondegenerate spectral gaps. This is an interesting and important problem that we do not address in this work.

\subsection{Properties of Gibbs states}
The Gibbs state of $H$ at inverse temperature $\beta$ is denoted by
\begin{equation}
    g_\beta=\frac{e^{-\beta H}}{\tr(e^{-\beta H})}.
\end{equation}
As mentioned in the main text, we always use units where the Boltzmann constant is $k_B = 1$.
Without loss of generality, we will only consider nonnegative temperatures $\beta\geq 0$ (a change on the overall sign of the Hamiltonian always ensures this). Below, we introduce several known properties satisfied by Gibbs states under certain conditions that we use throughout our work.
\subsubsection{Maximum entropy principle}
The defining property of the Gibbs state is that it uniquely maximizes the von Neumann entropy $S(\rho)=-\tr(\rho \log \rho)$ among all states sharing the same energy expectation value $E_\beta=\tr(g_\beta H)$, i.e.,
\begin{equation}
\label{eq:maxthermalentropy}
    g_\beta =\mathop{\textrm{argmax}}_{\rho: \tr(\rho H)=E_\beta} S(\rho).
\end{equation}
This can be shown by minimizing the relative entropy $D(\rho || g_\beta)=-S(\rho)+\beta\tr(H \rho) + \log(\tr(e^{-\beta H}))$.

\subsubsection{Finite correlation length}
The correlation length of a state denotes the effective distance over which correlations, whether classical or quantum, can persist. In many systems, particularly those at high temperatures, the Gibbs state exhibits a finite correlation length, defined as follows.
\begin{definition}[Finite correlation length]
\label{def:finitecorrlength}
    Let $\rho$ be a state.  We define the \emph{correlations between $A$ and $B$} by
    \begin{equation}
        \mathrm{Corr}_\rho(A,B)\coloneqq\max_{O_A,O_B} \frac{\abs{\tr(\rho O_A\otimes O_B)-\tr(\rho O_A)\tr(\rho O_B)}}{\norm{O_A}_{\mathrm{op}} \norm{O_B}_{\mathrm{op}}}
    \end{equation}
    where $O_A$ $(O_B)$ runs over all operators with support contained in region $A$ $(B)$.  We say $\rho$ has \emph{finite correlation length} $\xi>0$ if for any regions $A$ and $B$, 
$$\mathrm{Corr}_\rho(A,B) \leq f_{\mathrm{corr}}(a) e^{-d(A,B)/\xi},$$
where $f_{\mathrm{corr}}(a)$ is a fixed polynomial function, evaluated at $a=\min\{\abs{\partial A},\abs{\partial B}\}$, with $\abs{\partial A}, (\abs{\partial B})$ the size of the boundary of $A$ ($B$), i.e., the number of pairs of contiguous sites  such that one site is inside the region and the other outside. The distance between $A$ and $B$ is given by $$d(A,B)=\min\{\norm{\bm x-\bm y}_1\,|\,\bm x\in A, \bm y \in B\}.$$ 
\end{definition}

In one-dimensional systems, Gibbs states of TI local Hamiltonians have finite correlation length at nonzero temperatures, with $f_{\mathrm{corr}}(a)=1$ independent of $a$ \cite{Araki1969,Bluhm2022}. For higher dimensions, Gibbs states are guaranteed to have finite correlation length above a certain temperature, as shown by the following result.
\begin{lemma}[Theorem 2 of \cite{Kliesch2014}: High-temperature Gibbs states have finite correlation length]
\label{lemma:hightempgibbshavefinitecorr}
    Let $g_\beta$ be the thermal state at inverse temperature $\beta$ of a Hamiltonian defined over a collection of spins.
    There exist an inverse temperature $\beta_*^{(\mathrm{corr})}>0$   such that for every $\beta<\beta_*^{(\mathrm{corr})}$, there is a constant $\xi'$ such that the Gibbs state $g_\beta$ satisfies  $\mathrm{Corr}_\rho(A,B) \leq 4\,ac\,e^{-d_H(A,B)/\xi'}$
    for any regions $A,B$ with $d_H(A,B)\geq L_0(a)\coloneqq \xi' \abs{\log(ac)}$,
    where $a=\min\{\abs{\partial^{(H)} A},\abs{\partial^{(H)} B}\}$ and $c=(\log(3)(1-e^{-1/\xi'}))^{-1}$.
\end{lemma}
In Lemma~\ref{lemma:hightempgibbshavefinitecorr}, the  geometry is determined by the interaction terms of the general Hamiltonian, which subsequently determines the distance between regions $d_H(A,B)$ and their boundaries $\partial^{(H)}A,\partial^{(H)}B$. Specifically, Lemma~\ref{lemma:hightempgibbshavefinitecorr} considers the {\it interaction hypergraph} of $H$,  whose hyperedges are the supports of each of the interactions terms which make up $H$. The distance between two points in the interaction graph is the shortest path of hyperedges that connect them, and the boundary of a region consists of all hyperedges that contain a site inside the region and a site outside. 

In our setting, Lemma~\ref{lemma:hightempgibbshavefinitecorr} states that the Gibbs state has finite correlation length for $\beta<\beta_*^{(\mathrm{corr})}$, given by  \begin{equation}
\label{eq:crit-temp}
    \beta_*^{(\mathrm{corr})}=\frac{1}{2\max_i\norm{h_i}_{\mathrm{op}}}\log(1+\sqrt{1+4/\alpha}),
\end{equation} with $\alpha\geq((2R+1)^D-1)e$ \cite{Kliesch2014}, where $R$ is the side length of a hypercubical region  large enough to fit the (translated) support of any of the interaction terms $h_i$ appearing in the Hamiltonian. The  distance in the interaction hypergraph is related to the distance on the lattice by $d_H(A,B)\geq  d(A,B)/k$.
The extra factor $k$ arises because the support of $h_i$ can have at most diameter $k$ in the lattice. 
Similarly $|\partial^{(H)}A|\leq b_1 |\partial A|$, (with $b_1$ an unimportant constant which depends on $k$ and $D$). Because of the relations above, as long as $\beta<\beta_*^{(\mathrm{corr})}$, the Gibbs state has finite correlation length $\xi\leq \xi'/(D k)$, with $f_{\mathrm{corr}}(a)=b_1  4c^2 a^2$. The extra factor $(c a)$ in $f_{\mathrm{corr}}$ guarantees that the correlations are bounded for any distance $d(A,B)$, as we demand in Definition~\ref{def:finitecorrlength}, even if this distance is smaller than $L_0$. 
\subsubsection{Separability at high temperature}

Recently, Ref.~\cite{Bakshi2024}   showed that in any local qubit Hamiltonian, there is a constant temperature above which the Gibbs state is separable: it can be written exactly as a mixture of product states. 
\begin{lemma}[Theorem 1.5 of \cite{Bakshi2024}: High-temperature Gibbs states are separable]
\label{lemma:hightempareseparable} Consider a Hamiltonian consisting of Pauli-string interactions on qubits\footnote{Any local Hamiltonian can be written in terms of Pauli-string interactions.}, where each qubit appears in at most $\mathfrak{d}$ terms and each term contains at most $\mathfrak{K}$ qubits. 
For $|\beta|<\beta_*^{(\mathrm{sep})}\coloneq 1/(\gamma\mathfrak{d}\mathfrak{K}^2)$ with a fixed universal constant $\gamma$, the corresponding Gibbs state can be expressed as a mixture 
\begin{equation}
    g_\beta=\sum_{\ket{\psi}\in \mathcal{S}^{\otimes N}} p_\psi \dyad{\psi},
\end{equation} of product states in $\mathcal{S}^{\otimes N}$, where $\mathcal{S}=\{\ket{0},\ket{1},\ket{+},\ket{-},\ket{+ i},\ket{- i}\}$.
\end{lemma}
The states in $\mathcal{S}^{\otimes N}$ are product stabilizer states, i.e., $\ket{\pm}$, $\ket{\pm i}$ and $\ket{0/1}$ are the eigenstates of Pauli $X$, $Y$, and $Z$, respectively. The local terms of the Hamiltonian in Lemma~\ref{lemma:hightempareseparable} are assumed to have operator norm bounded by unity, which in our setting can always be achieved by dividing $H$ by $\max_i\norm{h_i}\leq O(1)$, thus effectively rescaling $\beta_*^{\mathrm{(sep)}}=1/(\max_i\norm{h_i}\gamma\mathfrak{d}\mathfrak{K}^2)$. The proof of Lemma~\ref{lemma:hightempareseparable} is constructive: Ref.~\cite{Bakshi2024} provides an explicit algorithm to efficiently sample from the distribution $p_\psi$. 
\subsubsection{Small subsystem purity} 
We will be studying the reduced Gibbs state $g_{\beta,A}=\tr_{\bar{A}}(g_\beta)$ within a region $A$. Intuitively, we expect the entropy of such state to be extensive, i.e., proportional to the number of qubits $N_A$ in $A$. Below we prove this intuition  at high temperature, phrased in terms of the subsystem purity being exponentially small in $N_A$.

\begin{prop}
\label{prop:bound_on_local_purity}
    Let $H$ be a local Hamiltonian defined over a $D$-dimensional periodic lattice and $A$ a region with $N_A$ qubits.  For $\beta\leq \beta_*^{\mathrm{(\mathrm{sep})}}$, the subsystem purity of the Gibbs state is bounded as

\begin{equation}
\label{eq:bound_on_purity}
       \tr((g_{\beta,A})^2)\leq e^{-\Omega(N_A)}.
    \end{equation} 
    \end{prop}
\begin{proof}
    We utilize the decomposition of the Gibbs state into stabilizer product states of Ref.~\cite{Bakshi2024}, given by Lemma~\ref{lemma:hightempareseparable}.
    We leverage following property of this probability distribution:
    \begin{lemma}
    \label{lemm:supportedeverywhere}
    Consider the distribution on $\mathcal{S}^{\otimes N}$ given by Lemma~\ref{lemma:hightempareseparable}.
       There exists a positive constant $c>0$ such that, for any $\ket{s}\in \mathcal{S}$, any site ${\bm j}\in \Lambda$, and any geometrically local $B\subseteq \Lambda\setminus \{ \bm j\}$, conditioned on all the sites in $B$ being on an arbitrary stabilizer product state $\ket{\psi_{B}}\in\mathcal{S}^{\otimes\abs{B}}$, the probability of $\ket{\psi_{\bm j}}=\ket{s}$   is lower bounded by $c$:
       \begin{equation}      \mathbb{P}\big[\ket{\psi_{\bm j}}=\ket{s}\, \big| \,\,|\psi_{B}\rangle\big]\geq c.
       \end{equation}
    \end{lemma}
\noindent Note that Eq.~\eqref{lemm:supportedeverywhere} is stronger than just saying that the marginal probability is lower bounded $\mathbb{P}\big[\ket{\psi_{\bm j}}=\ket{s}\, \big]\geq c$.  Lemma \ref{lemm:supportedeverywhere} states that even when knowing  the state of other sites, we can still expect any of the six outcomes for site ${\bm j}$ may occur with probability at least $c$. Lemma~\ref{lemm:supportedeverywhere} follows straightforwardly from analyzing the sampling algorithm of Ref.~\cite{Bakshi2024}, where we can see that any of the six outcomes may appear with constant probability as each qubit is {\it pinned}. Specifically, Algorithm~6.3 of Ref.~\cite{Bakshi2024} outputs $\mathds{1}/2$ as the state of the qubit with constant probability~\cite{Huang2024RPS}. 

    To apply Lemma~\ref{lemm:supportedeverywhere}, we write 
    \begin{equation}
    \tr((g_{\beta,A})^2)=\sum_{\ket{\phi},\ket{\psi}\in \mathcal{S}^{\otimes N}} p_\psi p_\phi \abs{\braket{\psi_A}{\phi_A}}^2\leq \sum_{\ket{\phi}\in \mathcal{S}^{\otimes N}} p_\phi \mathbb{P}_{\psi}[\forall {\bm j}\in A: \ket{\psi_{\bm j}}\neq \ket{\phi_{\bm j}}^\perp],
    \end{equation}
    where the probability $\mathbb{P}_\psi$ is taken with respect to the distribution $p_\psi$, and $\ket{\phi_{\bm j}}^\perp$ denotes the single-site state perpendicular to $\ket{\phi_{\bm j}}$, i.e., $\ket{0}^\perp = \ket{1}$, $\ket{+}^\perp = \ket{-}$, etc.

    By applying the chain rule for probability, we can use Lemma~\ref{lemm:supportedeverywhere} on each site in $ A=\{\bm a_1,\dots,\bm a_{N_A}\}$, conditioning on the {\it previous} sites:
    \begin{equation*}
        \mathbb{P}_{\psi}[\forall {\bm j}\in A: \ket{\psi_{\bm j}}\neq \ket{\phi_{\bm j}}^\perp]=\prod_{ k=1}^{N_A}\mathbb{P}_{\psi}\big[ \ket{\psi_{\bm a_k}}\neq \ket{\phi_{\bm a_k}}^\perp\,\big| \,|\,\psi_{A\setminus\{{\bm a}_1,\dots,{\bm a}_k\}}\rangle\big]\leq (1-c)^{N_A}.\qedhere
    \end{equation*}
\end{proof}
\subsection{Depth complexity and shallow-complexity states}
We prove thermalization on systems initially prepared on some {\it physical} quantum state. To precisely define what we mean by physical, we use the notion of depth complexity.
\begin{figure}[bt]
    \centering
\includegraphics[width=1\columnwidth]{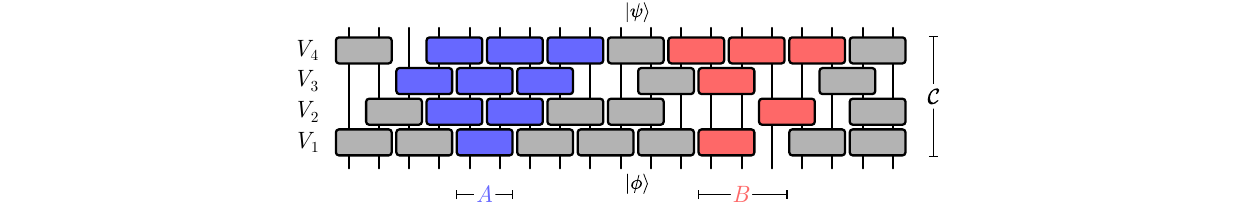}
    \caption{State $\ket{\psi}$ with complexity $\mathcal{C}=4$ in 1D is generated from a product state $\ket{\phi}$ by $4$ circuit layers $V_1$-$V_4$. Each box represents a unitary gate. Blue and red boxes mark the light cone of two regions $A$ and $B$, respectively.}
    \label{fig:S01}
\end{figure}

\begin{definition}[Shallow complexity state]
    Given any state $\ket{\psi}$, its \emph{(depth) complexity} $\mathcal{C}$ is the minimum number of circuit layers that are necessary to transform a product state $\ket{\phi}$ into $\ket{\psi}$,
    \begin{equation}
\ket{\psi}=V_{\mathcal{C}}\cdots V_2V_1\ket{\phi}.
\end{equation}
where a layer is a unitary $V_j\in U(2^N)$ made up of products of geometrically local $2$-qubit gates with disjoint support (Supp. Fig.~\ref{fig:S01}).
We say that $\ket{\psi}$ has \emph{shallow complexity} if its depth complexity is subpolynomially growing with $N$, meaning that $\mathcal{C}\leq O(N^\nu)$  for any $\nu>0$. 

\end{definition}

Note that product states themselves have shallow complexity, with trivial $\mathcal{C}=0$, and increasing $\mathcal{C}$ increases the possible ammount of entanglement in the state. Crucially,
  we may upper bound the entanglement and the long-range correlations of a state via its complexity.
\begin{lemma}[Bound on entanglement and correlations from depth complexity]
\label{lemm:propsofconstdepthcomplexity}
    Let $\ket{\psi}$ have depth complexity $\mathcal{C}$. Then
    \begin{enumerate}
        \item[(i)] For any region $A \subseteq \Lambda$, the max entropy  $S_0(\psi_A)=\log(\mathrm{rank}(\psi_A))$ is upper bounded by the number of sites in $A$ which are at most a distance $\mathcal{C}$ away from the boundary, i.e. $S_0(\psi_A)\leq \log(2){\abs{\partial_{\mathcal C} A} }$, with  $\partial_{\mathcal{C}} A =\{\bm a\in A\,|\,\exists\bm b\in \bar{A}:\, \norm{\bm a -\bm b}\leq \mathcal{C}\}$.  
        \item[(ii)] The state $\ket{\psi}$ does not have correlations beyond (twice) its depth complexity. For any regions $A,B\subseteq \Lambda$ with $d(A,B)>2\mathcal{C}$,
        \begin{equation}
            \mathrm{Corr}_\psi(A,B) =0.
        \end{equation}
    \end{enumerate}
\end{lemma}
\begin{proof} 
For part {\it (i)} note that $\mathrm{rank}(\psi_A)\leq 2^m$, where $m$ counts  number of qubits in $A$ which are entangled with qubits in $\bar{A}$.  Because each layer is composed of local $2$-qubit gates, for a qubit in $A$ to be entangled with a qubit outside, it must be at most a distance $\mathcal{C}$ away in from the boundary, so $\mathrm{rank}(\psi_A)\leq 2^{\abs{\partial_{\mathcal C} A} }$, as desired.

For part {\it (ii)}, note that $\bra{\psi}O_A\otimes O_B\ket{\psi}=\bra{\phi}U^\dagger (O_A\otimes O_B)U\ket{\phi}$, where $U=V_{\mathcal C}\cdots V_2 V_1$ is the circuit that maps the product state $\ket{\phi}$ into $\ket{\psi}$. If $d(A,B)\leq 2\mathcal{C}$, then the support of $U^\dagger O_A U$ does not intersect that of $U^\dagger O_A U$ ---it is said that $A$ lies outside of the {\it light cone} of $B$, as illustrated in  Supp. Fig.~\ref{fig:S01}--- so we may factorize $\bra{\phi}U^\dagger O_A \otimes O_B U\ket{\phi}=\bra{\phi}U^\dagger O_A U\dyad{\phi}U ^\dagger O_B U\ket{\phi}$, or $\bra{\psi}O_A \otimes O_B \ket{\psi}=\expval {O_A}{\psi}\expval{O_B}{\psi}$.
\end{proof}

If the state $\ket{\psi}$ has constant complexity, Lemma~\ref{lemm:propsofconstdepthcomplexity}~(i) implies an area law for the entanglement of  $S(\psi_A)\leq O(\abs{\partial A})$.
\section{The EDGE theorem}
\label{sec:EDGE}
We are now in position to present our first result. We consider 
energy dispersed Gibbs ensembles (EDGEs) (see Definition~\maintextref{1} from the main text) and prove our EDGE theorem:
\begin{theorem}[EDGE Theorem \maintextref{2}, extended statement]
\label{th:EDGE}
Consider any translation invariant, geometrically local Hamiltonian for $N$ qubits in $D$-dimensional lattice with nondegenerate spectral gaps.
    Let $\mathcal{E}_\beta$ be an EDGE and assume that the Gibbs state $g_\beta$ has finite correlation length.
    Then, for any geometrically local region $A$ with a constant number of qubits $N_A$,  \begin{align}
        \E_{\psi}& \E_t \left[\norm{g_\beta-\psi(t)}_A \right] \leq O(N^{-\gamma})+2\varepsilon_{\mathrm{GE}}(N)+{2^{N_A}} \varepsilon_{\mathrm{ED}}(N)^{1/2}, \label{eq:boundav}
    \end{align}
for any $\gamma<1/(D+1)$. Furthermore, the statement holds for any $\gamma<1/2$ (independent of $D$) as long as the temperature is above a threshold $\beta<\beta_*^{(\mathrm{con})}$ given by $(\beta_{*}^{(\mathrm{con})})^{-1}= 8e^3\times \max\{N_A^2,\mathfrak{d}\mathfrak{K}\max_i\norm{h_i}\}$, where, in the Hamiltonian, each qubit appears in at most $\mathfrak{d}$ terms and each term $h_i$ contains at most $\mathfrak{K}$ qubits.
\end{theorem}

The rest of this section is devoted to proving Theorem~\ref{th:EDGE}. Our proof is inspired by Ref.~\cite{Huang2019}, which studies 1D systems at infinite temperature. We will show that a typical state $\psi$ in the EDGE satisfies two properties:
\begin{enumerate}
    \item[(a)] Equivalence of ensembles between the temporal $\rho_\infty=\E_t[\psi(t)]$ and the thermal ensemble $g_\beta$: $$\E_{\psi}\norm{\rho_\infty - g_\beta}_A\leq  O(N^{-\gamma})+2\varepsilon_{\mathrm{GE}}(N),$$
    \item[(b)] Equilibration:
    $$\E_{\psi}\E_t[\norm{\rho_\infty-\psi(t)}_A]\leq   {2^{N_A}} \varepsilon_{\mathrm{ED}}(N)^{1/2}.$$
\end{enumerate}
Taken together, these two properties give us the desired quantum thermalization in Eq~\eqref{eq:boundav}.
 The equivalence of ensembles will follow from property (a) of an EDGE and a variant of the weak ETH \cite{Biroli2010,Mori2016,Brandao2019,Alhambra2020}. We will explain this first. After, we will explain how  equilibration follows from property (b) of an EDGE and the seminal results on quantum equilibration from Ref.~\cite{Linden2009}.

 Before proceeding, let us make a comment on the error term $O(N^{-\gamma})$ in Eq.~\eqref{eq:boundav}, which is the finite-size error term for the equivalence of ensembles coming from the weak ETH.  As we will see below, the scaling of $\gamma$ depends on known concentration bounds for the expectation value of extensive observables. The regular ETH ansatz predicts a finite-size error for thermalization of $O(N^{-1})$ ($\gamma=1$) \cite{Srednicki1996,Huang2024}, and thus even the best scaling $\gamma=1/2-\delta$ proven in 1D and in any dimension at high temperature is not expected to be optimal.  Rigorously proving this better bound under minimal assumptions is an important further direction.
 
\subsection{Equivalence of the temporal and thermal ensembles}
The weak ETH guarantees that with high likelihood, the expectation value of local observables over the energy eigenstates of a translation invariant system  is close to the expectation at equilibrium, with probability exponentially close to 1. It was originally derived in Refs.~\cite{Biroli2010,Mori2016}, and then further improved in different settings~\cite{Brandao2019,Alhambra2020,Kuwahara2020}. We present here a modified version which is favorable for our purposes.

\begin{prop}[A variant of the weak ETH]
\label{prop:weakETH}
     Let $H$ be a TI, (geometrically) local Hamiltonian on a $D$-dimensional periodic lattice with $N$ qubits.
    Let $\rho=\sum_{j=1}^{2^N}p_j\dyad{E_j}$ be a diagonal state in the common eigenbasis $\{\ket{E_j}\}_j$ of $H$ and the translation operator, that has finite correlation length. Consider a geometrically local region $A$.  Then \begin{equation}
        \label{eq:weakETH}
        \mathbb{P}_{j\sim p_j}\Big[\big\|\dyad{E_j} - \rho\,\big \|_A\geq\varepsilon \Big]\leq e^{-\Omega(N^{\nu}\varepsilon) },
    \end{equation} 
    for $\nu= 1/(D+1)$ and $\varepsilon\geq \Omega(N^{-\gamma})$ with $\gamma<\nu$. Furthermore, if the state $\rho$ is a Gibbs state of $H$, $\rho=g_\beta$ at sufficiently high temperature $\beta< \beta_{*}^{(\mathrm{con})}$ (defined in Theorem \ref{th:EDGE}), then the statement holds for $\nu=1/2$, regardless of $D$.
    \begin{proof}
    
        We consider an arbitrary observable supported on region $A$ with operator norm $\norm{O_A}_{\textrm{op}}\leq 1$, over which we will optimize later to obtain the local trace norm. Following  Refs.~\cite{Mori2016,Brandao2019,Alhambra2020}, we first show that
        \begin{equation}
            \label{eq:proboundweth}
            \mathbb{P}_{j\sim p_j}\Big[\expval{O_A}{E_j} - \tr(O_A\rho)\geq\varepsilon \Big]\leq e^{-\Omega(N^{\nu}\varepsilon) }.
        \end{equation} 
        
        The first step is to utilize the translation invariance of the Hamiltonian to turn the expectation values of $O_A$ into expectation values of an extensive observable  $\bar{O}=\sum_{\bm n\in \Lambda}\mathbb{T}^{\bm n} O_A \mathbb{T}^{-\bm n} - N\tr(\rho O_A)$, such that $\langle E_j |O_A |E_j\rangle - \tr(\rho  O_A)=\frac{1}{N} \expval{\overline{O}}{E_j}$. Then, for any $\tau\geq 0$, 
        \begin{align}
    \mathbb{P}_{j}\Big[{\langle E_j|O_A |E_j\rangle - \tr(\rho  O_A)}\geq\varepsilon \Big]&=\mathbb{P}_j\Big[e^{\tau {\expval{\overline{O}}{E_j}}}\geq e^{N\tau\varepsilon} \Big]\\&\leq 
         e^{-N\tau \varepsilon}\sum_{j=1}^{2^N} p_je^{\tau{\expval{\overline{O}}{E_j}}}\\&\leq  e^{-N\tau\varepsilon} \sum_{j=1}^{2^N} p_j\expval{e^{\tau{\overline{O}}}}{E_j}= e^{-N\tau \varepsilon}\tr(\rho \,e^{\tau \overline{O}}),
    \label{eq:sumtotraceonabsvalue}
    \end{align}
    where the second inequality follows from Jensen's inequality.
    
    The purpose of recasting the inequality in this form is to utilize known concentration bounds for the expectation value of extensive local observables. Equation~\eqref{eq:proboundweth} is proven for $\nu=1/(D+1)$ in Ref.~\cite{Alhambra2020} using concentration bounds from Ref.~\cite{Anshu2016}. Here, we prove that it further holds for $\nu=1/2$ for any $D$ whenever $\rho=g_\beta$  at high enough temperature, by leveraging the concentration bound of Ref.~\cite{Kuwahara2020}. Specifically, we apply the following result:
    \begin{lemma}[Theorem 1 of Ref.~\cite{Kuwahara2020}]\label{lemm:largediationbound} Let $\mathcal{F}=\sum_{\abs{X}\leq k} f_X$ and $H=\sum_{\abs{X}\leq k} h_X$  with $f_X,h_X$ $k$-body operators with bounded operator norm $\sum_{X:{X\ni \bm x}} \norm{f_X}_{\mathrm{op}},\sum_{X:{X\ni \bm x}} \norm{h_X}_{\mathrm{op}}\leq g$ for any site $\bm x$. Then, if the inverse temperature satisfies $\beta<\beta_c\coloneqq(8e^3gk)^{-1}$, the Gibbs state $\rho$ satisfies the following inequality:
    \begin{equation}
        \log\tr(e^{-\tau \mathcal{F}}\rho)\leq -\tau \tr(\mathcal{F}\rho) +\frac{\tau^2B}{\beta_c-\beta-\abs{\tau}},
    \end{equation}
    for $\abs{\tau}<\beta_c-\beta$ and $B\coloneqq\sum_{\abs{X}\leq k} \norm{f_X}_{\mathrm{op}}$.
    \end{lemma}
    In our particular scenario, $\mathcal{F}=\sum_{\bm n\in \Lambda}\mathbb{T}^{\bm n}O_A\mathbb{T}^{-\bm n}$,  $B=N\norm{O_A}_{\mathrm{op}}$$ $, so if we choose $\tau=N^{-1/2}$  we obtain $\tau^2B/(\beta_c -\beta -\abs{\tau}))\leq O(1)$ and

    \begin{equation}
        \mathbb{P}_{j}\Big[{\langle E_j|O_A |E_j\rangle - \tr(\rho  O_A)}\geq \varepsilon\Big]\leq e^{-N\tau \varepsilon}\tr(\rho \,e^{\tau \overline{O}})\leq e^{-\Omega(N^{1/2}\varepsilon)},
    \end{equation}
     which completes the proof of Eq.~\eqref{eq:proboundweth} in the high temperature case. Note that the critical temperature has dependence on both the locality of the Hamiltonian and $N_A^2$ (because the terms of $\mathcal{F}$ are all the translates of $O_A$, which is a $N_A$-body operator).

    Now, to finally obtain Eq.~\eqref{eq:weakETH}, we need to maximize Eq.~\eqref{eq:proboundweth} over $O_A$. Note, importantly, that we have to perform the maximization separately for each $j$. For this, we consider a finite set of observables and perform a union bound \cite{Brandao2019}. Specifically, let $\mathcal{N}$ be a $(\varepsilon/3)$-net of observables supported on $A$, which means that for any $O_A$ with  norm bounded by $1$, there is some $O_A'\in\mathcal{N}$ such that $\norm{O_A-O_A'}_{\mathrm{op}}\leq \varepsilon/3$. $\mathcal{N}$ can be taken to have at most $\abs{\mathcal{N}}\leq O((1/\varepsilon) ^{\alpha})$ elements, where $\alpha$ is a constant which only depends on $N_A$. Then, for each $j$, there is a $O_A^{(j)}\in \mathcal{N}$ such that $\big\|\dyad{E_j} - \rho\,\big \|_A\leq \expval{O_A^{(j)}}{E_j} - \mathrm{tr}(O_A^{(j)}\rho)+\tfrac{2}{3}\varepsilon$, and so
    \begin{align}
\mathbb{P}_{j}\Big[\big\|\dyad{E_j} - \rho\,\big \|_A\geq\varepsilon \Big]&\leq \mathbb{P}_{j}\Big[\exists O_A^{(j)}\in \mathcal{N}\colon \expval{O_A^{(j)}}{E_j} - \mathrm{tr}(O_A^{(j)}\rho)\geq\varepsilon/3 \Big] \\&\leq \sum_{O_A\in\mathcal{N}}\mathbb{P}_{j}\Big[\expval{O_A}{E_j} - \mathrm{tr}(O_A\rho)\geq\varepsilon/3 \Big] \leq O(\varepsilon^{-\alpha}) e^{-\Omega(N^\nu \varepsilon)}\leq e^{-\Omega(N^\nu \varepsilon)},
\end{align}  
where in the last inequality we used that $\varepsilon\geq \Omega(N^{-\gamma})$ with $\gamma<\nu$.
    \end{proof}
\end{prop}
From Proposition~\ref{prop:weakETH} we can prove Proposition~\maintextref{1} of the main text, stated here in slightly more general form.
\begin{corollary}[Generalization of Proposition~\maintextref{1}]
\label{cor:similarformofweakETH}
     Under the conditions of Proposition~\ref{prop:weakETH}, for any $\gamma<\nu$, \begin{equation}
        \label{eq:weakerETH}
        \sum_{j=1}^{2^N} p_j \big\|\dyad{E_j} - \rho\,\big \|_A\leq O(N^{-\gamma})
    \end{equation} 

    \begin{proof}
    From Proposition~\ref{prop:weakETH} we obtain
    \begin{align}
                \sum_{j=1}^{2^N} p_j \big\|\dyad{E_j} - \rho\,\big \|_A\leq \varepsilon + 2e^{-\Omega(N^\nu\varepsilon)},
    \end{align}
    and if we pick $\varepsilon=N^{-\gamma}$, we get $\varepsilon + 2e^{-\Omega(N^\nu\varepsilon)}\leq O(N^{-\gamma})$.
        \end{proof}
    
\end{corollary}

\begin{prop}[Equivalence of temporal and thermal ensembles]
\label{prop:EDGEeqofens}

Under the assumptions of the EDGE Theorem \ref{th:EDGE}, 
\begin{align}
\label{eq:EDGEeqofens}
    \E_{\psi\in  \mathcal{E}_\beta} [\norm{\rho_\infty-g_\beta}_A]\leq O(N^{-\gamma})+2\varepsilon_{\mathrm{GE}}(N),
\end{align} 
where $\rho_\infty=\E_t[\psi(t)]$.
\end{prop}
\begin{proof}
    
By the no-degeneracy condition, the infinite-time-average of any state $\rho_{\infty}$ is the diagonal ensemble $\rho_\infty=\sum_j \abs{\braket{\psi}{E_j}}^2 \dyad{E_j}$.  Then
    \begin{align*}
    \E_{\psi\in \mathcal{E}_\beta}[\norm{\rho_\infty(\psi) -g_\beta }_A]&=\E_{\psi\in \mathcal{E}_\beta}\bigg[\Big\|\sum_j \big (\dyad{E_j} -g_\beta\big)  \abs{\braket{\psi}{E_j}}^2\Big\|_A\bigg]\\
    &\leq {\sum_j \norm{\dyad{E_j} -g_\beta}_A \E_{\psi\in \mathcal{E}_\beta}\Big[\abs{\braket{\psi}{E_j}}^2\Big]}\\
    &= {\sum_j  \norm{\dyad{E_j} -g_\beta}_A\bra{E_j}\E_{\psi\in \mathcal{E}_\beta}[\psi ]\ket{E_j}}\\
    &\leq \sum_j  \norm{\dyad{E_j} -g_\beta}_A \left(\!\expval{g_\beta}{E_j}+2^{-N} \varepsilon_{\mathrm{GE}}(N)\right)
    \\
    &\leq \Big(\sum_j  \norm{\dyad{E_j} -g_\beta}_A \expval{g_\beta}{E_j}\Big)+2\varepsilon_{\mathrm{GE}}(N)\\&\leq O(N^{-\gamma})+2\varepsilon_{\mathrm{GE}}(N),
    \end{align*}

    where we used property (a) of the EDGE and Corollary~\ref{cor:similarformofweakETH}.
    \end{proof}

\subsection{Quantum equilibration}

Now we shift gears to prove equilibration.  We utilize the following result from Ref.~\cite{Linden2009}:

\begin{lemma}[Quantum equilibration, Theorem 1 of Ref.~\cite{Linden2009}]
    \label{lemm:quantumEq}
    Let $H$ be a Hamiltonian which has nondegenerate spectral gaps. For any state $\psi$ with time average $\rho_\infty=\E_t[\psi(t)]$, and subsystem $A$ of dimension $d_A$
    \begin{equation}
        \E_t [\norm{\rho_\infty-\rho(t)}_A ] \leq {d_A}\sqrt{\mathrm{IPR}}, 
    \end{equation}
    where $\mathrm{IPR}=\sum_j \abs{\braket{E_j}{\psi}}^4$ is the inverse participation ratio of $\ket{\psi}$.
\end{lemma}

\begin{prop}[EDGE equilibration]
\label{prop:EDGEeq}
    Under the assumptions of the EDGE Theorem \ref{th:EDGE},
\begin{align}\label{eq:EDGEequil}
    \E_{\psi\in  \mathcal{E}_\beta} [\E_t [\norm{\rho_\infty-\rho(t)}_A ] ]\leq {2^{N_A}} \varepsilon_{\mathrm{ED}}(N)^{1/2}
\end{align} 
\end{prop}
\begin{proof}
     Averaging over all states in the EDGE $\mathcal{E}_\beta$, using that $\E[\,\cdot\,]^2\leq \E[(\,\cdot\,)^2]$, and property (b) in Definition~\maintextref{1} of the main text, we obtain  
\begin{equation*}
    \E_{\psi\in  \mathcal{E}_\beta} [\E_t [\norm{\rho_\infty-\rho(t)}_A ] ]\leq \sqrt{\E_{\psi\in  \mathcal{E}_\beta} [\E_t [\norm{\rho_\infty-\rho(t)}_A ]^2 ]}\leq {2^{N_A}} \varepsilon_{\mathrm{ED}}(N)^{1/2}. \qedhere
\end{equation*}
\end{proof}

Combining Propositions~\ref{prop:EDGEeqofens} and~\ref{prop:EDGEeq}, by summing Eq.~\eqref{eq:EDGEeqofens} and Eq.~\eqref{eq:EDGEequil} and applying a triangle inequality, we complete the proof of the EDGE Theorem~\ref{th:EDGE}.

In the next subsection, we present a  generalization of the EDGE Theorem to finite time. We also present a generalization to generalized Gibbs states in Section~\ref{sec:generalized}.

\subsection{Finite-time thermalization}
We note that the EDGE theorem, as stated, does not say anything about the time it takes for systems to thermalize. However, using known finite-time bounds on equilibration \cite{Wilming2018}, we can easily reformulate the EDGE theorem to become a finite-time statement. For example, Eq.~(24) of Ref.~\cite{Short2012} states that, under the assumptions of Lemma~\ref{lemm:quantumEq},\footnote{This bound can be slightly improved  using \cite[Eq. (A9)]{Malabarba2014}. }
 \begin{equation}
 \label{eq:th3shortfinitetimeq}
    \E_{t\in[0,T]} [\norm{\rho_\infty - \psi(t)}_A]\leq d_A \sqrt{ \mathrm{IPR}\times\left(1 +\frac{8N}{G_{\min} T}\right)},
\end{equation}
where $\E_{t\in[0,T]}[\,\cdot\,]=\frac{1}{T}\int_0^T\dd t\,(\,\cdot\,)$ denotes a finite-time average and \begin{equation}
    G_{\min}=\min_{(l,m)\neq (i,j)\neq (m,l)}{\abs{(E_i  -  E_m) - (E_l -E_j)} }
\end{equation}  is the smallest {\it gap of gaps}.
This quantity is nonzero due to the nondegenerate gap condition, Eq.~\eqref{eq:nondegenerategapcondition}.  Using this result we obtain:

\begin{theorem}[Finite-time EDGE]
\label{th:finite-time-EDGE}
    Under the conditions of the EDGE Theorem~\ref{th:EDGE},
\begin{equation}
            \E_{\psi\in E_\beta} \E_{t\in[0,T]} \left[\norm{g_\beta-\psi(t)}_A \right] \leq O(N^{-\gamma})+2\varepsilon_{\mathrm{GE}}(N)+2^{N_A}\sqrt{\varepsilon_{\mathrm{GE}}(N)\left(1 +\frac{8N}{G_{\min} T}\right)}. \label{eq:boundavfinitetime}
\end{equation}
\end{theorem}
Theorem~\ref{th:finite-time-EDGE} guarantees that for any typical state in the ensemble, thermalization is achieved when $8N/G_{\min} T\sim 1$, i.e. after time $T_{\mathrm{therm}}\sim N/G_\mathrm{min}$. 

One can estimate how small $G_\mathrm{min}$ as a function of increasing system sizes with the following argument.
Let us consider $M\sim2^{4N}$ different gaps of gaps $\{\delta \Delta_ 1, \delta \Delta_ 2,\dots \delta \Delta_M \}$.
To simplify our analysis we model them as if they are independent random variables uniformly distributed within a range $[0,W]$. We can exactly compute the probability distribution for $\delta \Delta_\textrm{min} \equiv \textrm{min}\{\delta \Delta_i\}_{i=1}^{M} $ and evaluate the average $\overline{\delta \Delta_\textrm{min}}=W/(M+1)$. Thus, we should expect $G_\mathrm{min}\sim 2^{-4N}$. This scaling can be numerically verified in translation-invariant local Hamiltonians (not presented here). Hence, the bound $T_{\mathrm{therm}}$ is expected to grow as $\sim 2^{4N}$. An exponential scaling in the bound is inevitable, as there indeed exist translation-invariant systems that can thermalize only after exponentially long times~\cite{Balasubramanian2024}.

\section{Maximally entropic ensembles of shallow-complexity states at finite temperature}
\label{sec:MEE}
We present a more detailed definition of   maximally entropic ensembles (MEEs), and prove some of their properties. Let us begin by precisely stating what we mean by complexity and temperature associated to an ensemble.
\begin{definition}[Complexity of an ensemble]
    We say that an ensemble of pure states $\mathcal{E}_{\mathcal{C}}$ has \emph{complexity} $\mathcal{C}(N)$ if any state $\psi$ in the support of $\mathcal{E}_{\mathcal{C}}$ has complexity at most $\mathcal{C}(N)$, i.e. it can be prepared by a circuit of depth at most $\mathcal{C}(N)$ from a product state. 
\end{definition}

\begin{definition}[Temperature of an ensemble]
    We say that an ensemble $\mathcal{E}_{\beta}$ of pure states has \emph{temperature} $1/\beta$ if its average energy is matches with that of the thermal state
    \begin{equation}
        \E_{\psi\in \mathcal{E_\beta}}[\expval{H}{\psi}] =\tr(H g_\beta).
    \end{equation}
\end{definition}
We define MEEs, as follows.
\begin{definition}[Maximally entropic ensemble]
    An ensemble $\mathcal{E^*_{\beta,\mathcal{C}}}$
with temperature $1/\beta$ and complexity $\mathcal{C}$ is a \emph{maximally entropic ensemble} (MEE) if the von Neumann entropy $S(\rho)=-\tr(\rho \log(\rho))$ of its average state $\rho^*_{\beta,\mathcal{C}}=\E_{\psi\in \mathcal{E}^*_{\beta,\mathcal{C}}}[\psi]$ is maximal among all ensembles with the same temperature and complexity, i.e.
\begin{equation}
\mathcal{E}^*_{\beta,\mathcal{C}}\in \mathop{\mathrm{argmax}}_{\substack{\mathcal{E}_{\beta,\mathcal{C}}}} \,S(\rho_{\beta,\mathcal{C}}).
\end{equation}
\end{definition}

Importantly, note that MEEs are not unique. For example, all of the following ensembles form a MEE of product states (zero complexity) at infinite temperature $\mathcal{E}_{\beta=0,\mathcal{C}=0}$: the ensemble of all computational basis states, the ensemble of all stabilizer product states, and the ensemble of all product states (each qubit state Haar-distributed). This is simply because in all cases, the average state is the maximally entropic state $\mathds{1}/2^N$. Beyond product states, the ensemble of outputs of a random unitary circuit of some fixed shallow depth is also a MEE, for the same reason. At finite, but high enough temperature $\beta\leq \beta_*^{(\mathrm{sep})}$, the ensemble of stabilizer product states $\mathcal{E}_{\mathcal{C}=0,\beta}$ given by Lemma~\ref{lemma:hightempareseparable} is a MEE, because its average state is the Gibbs state, which has maximal entropy at fixed energy. One can also transform a MEE into a different MEE: given any quantum circuit $U$ with $\mathcal{C}$ layers, the ensemble $\mathcal{E}_{\mathcal{C},0}=\{U\ket{\psi}\,|\, \ket{\psi}\in\mathcal{E}_{\beta=0,\mathcal{C}=0}\}$ is a MEE at $\beta=0$ with complexity $\mathcal{C}$. Furthermore, if the circuit leaves the Gibbs state invariant $U^\dagger g_\beta U=g_\beta$, then $\mathcal{E}_{\mathcal{C},\beta}=\{U\ket{\psi}\,|\, \ket{\psi}\in\mathcal{E}_{\beta,\mathcal{C}=0}\}$ is a MEE of complexity $\mathcal{C}$ for any $\beta\leq \beta_*^{(\mathrm{sep})}$.

Our main result, Theorem~\maintextref{1}, guarantees the thermalization of any MEE of shallow complexity at sufficiently high temperature. This is a consequence of the EDGE theorem: below we prove that at high enough temperature any MEE is a Gibbs ensemble, and we further prove in Sec.~\ref{sec:boundonaverageIPR} that any Gibbs ensemble of shallow-complexity states is an EDGE.
\subsection{Maximally entropic ensembles are Gibbs ensembles at high temperature}
Here, we show that at high enough but finite temperatures, a MEE forms a Gibbs ensemble.
\begin{lemma}[Lemma~\maintextref{1} restatement]
\label{lemma:HT-MEE-GE}
For any geometrically local Hamiltonian and any choice of complexity $\mathcal{C}(N)$, there exists a constant threshold  $\beta_*^{(\mathcal{C})}\geq \beta_*^{(\mathrm{sep})}$ which is independent of $N$ such that any MEE  $\mathcal{E}_{\beta,\mathcal{C}}$ with complexity $\mathcal{C}$ and inverse temperature $|\beta|\leq \beta_*^{(\mathcal{C})}$ averages to the Gibbs state, $\E_{\psi}[\psi]=g_\beta$.
Furthermore, the threshold temperature does not increase as the complexity increases: $\beta_*^{(\mathcal{C})} \geq \beta_*^{(\mathcal{C'})}$ if $\mathcal{C}\geq \mathcal{C'}$.
\end{lemma}
\begin{proof}
Let us fix the complexity $\mathcal{C}(N)$. 
 Define $\beta_*^{(\mathcal{C})}>0$ to be the smallest inverse temperature such that the Gibbs state at any $\beta\leq \beta_*^{(\mathcal{C})}$ is exactly decomposable as some mixture of pure states of circuit complexity at most $\mathcal{C}$. Note that, by the separability of the Gibbs state stated in Lemma~\ref{lemma:hightempareseparable}, we know that $\beta_*^{(\mathcal{C})}\geq \beta_*^{(\mathrm{sep})}>0$. Furthermore, from this definition we immediately get the monotonicity $\beta_*^{(\mathcal{C})} \geq \beta_*^{(\mathcal{C'})}$ if $\mathcal{C}\geq \mathcal{C'}$.  
 
 Now we prove that any MEE $\mathcal{E}_{\beta,\mathcal{C}}$ with inverse temperature $\beta\leq \beta_*^{(\mathcal{C})}$ and complexity $\mathcal{C}$ is a Gibbs ensemble. Let $\mathcal{E}_{\beta,\mathcal{C}}^*=\{(p_\psi,\dyad{\psi})\}$ be an ensemble of pure states of complexity at most $\mathcal{C}$ such that $\E_\psi[\psi]=g_\beta$ (it exists by how we defined $\beta_*^{(\mathcal{C})}$). The ensemble $\mathcal{E}^*_{\beta,\mathcal{C}}$  is a MEE because the thermal state $g_\beta$ precisely maximizes the von Neumann entropy $S(\rho)=-\tr(\rho \log \rho)$ with constrained energy [Eq.~\eqref{eq:maxthermalentropy}]. Furthermore, $g_\beta$ uniquely satisfies this property, so the average state of any other MEE at the same temperature must also be the Gibbs state, regardless of the complexity. 
    \end{proof}
In Sec.~\ref{sec:boundonaverageIPR} we prove that, further assuming shallow-complexity, one can show that MEEs are EDGEs, which together with our EDGE theorem completes our proof of thermalization. Before this, we present a brief discussion about other ensembles of pure states for which the statement of thermalization is interesting, which are different from MEEs.

\subsection{Non-MEE ensembles of product states}
 We briefly discuss other natural finite-temperature ensembles of pure states which  are not obtained through our MEE construction. For simplicity, we restrict this discussion to product states. Given any probability density $p(\psi)$ in the manifold of product states, to restrict to finite temperature, one requires an energy constraint $\int d\psi p(\psi) \expval{H}{\psi} = E_\beta$. When we define MEEs, we implicitly select $p$ to be whatever distribution that maximizes the von Neumann entropy of the average state. However, one can select such $p$ explicitly instead. Two natural choices are a {\it microcanonical product state ensemble}~\cite{Huang2024RPS},
 \begin{equation}
     \label{eq:productstatemic}
     p_{(\mathrm{mic})}(\psi)=\begin{cases}
     \frac{1}{M}&\mathrm{ if  } 
 \quad\abs{\expval{H}{\psi} - \tr(g_\beta H)\,}\leq \Delta(N),\\
     0&\mathrm{otherwise.}
     \end{cases}
 \end{equation}
 for some $\Delta(N)=o(N)$ or a {\it canonical product state ensemble} via a Boltzmann factor
  \begin{equation}
     p_{(\mathrm{can})}(\psi)=\frac{1}{Z'}e^{-\beta'\expval{H}{\psi}}
 \end{equation}
where $M$ and $Z'$ are chosen to ensure normalization.

It is a natural question to ask whether these ensembles typically thermalize. We note, however, that one has to be careful in interpreting them as ensembles of statistical mechanics. One is effectively treating each state 
 $\psi$ as a distinct state with a well-defined energy $\expval{H}{\psi}$,  ignoring its quantum mechanical nature. Even if $p_{(\textrm{mic})}(\psi)>0$, the state $\psi$ might still have overlap with energy eigenstates outside of the microcanonical window, because it is only the expectation value that is constrained. On the other hand the Boltzmann factor $\beta'$ in $p_{(\textrm{can})}$ will be in general different from the physical inverse temperature $\beta$, due to an over-counting in entropy,  stemming from ignoring the partial indistinguishability of non-orthogonal states. We do not expect that these ensembles are Gibbs ensembles, in the sense of the EDGE theorem. Proving their thermalization is nevertheless an interesting question that we leave open. In this direction, Ref.~\cite{Huang2024RPS} proved the equilibration (although not the thermalization) of typical states in  the microcanonical product-state ensemble of Eq.~\eqref{eq:productstatemic}.

\section{Bound on average IPR}
\label{sec:boundonaverageIPR}
We discussed in Sec.~\ref{sec:EDGE} that MEEs at high temperature satisfy property (a) of an EDGE, with $\varepsilon_{\mathrm{GE}}(N)=0$. Here, we prove that, as long as the MEE has shallow complexity, it also satisfies property (b) of an EDGE with  $\varepsilon_{\mathrm{ED}}(N)\leq O(N^{-1+\delta})$, which completes the proof of Theorem~\maintextref{1}.

We prove Theorem~\maintextref{3} of the main text:
\begin{theorem}[Bound on average IPR, Theorem~\maintextref{3}  restatement]
\label{th:boundonIPR}
    In a periodic lattice with $N$ qubits, consider an ensemble $\mathcal{E}$ of pure states with shallow complexity $\mathcal{C}(N)$  whose average state $\rho=\E_{\psi\in \mathcal{E}}[\psi]$ satisfies the following three properties:
     \begin{enumerate}
         \item[(i)] Translation invariance, $\forall \bm n\in\Lambda: \mathbb{T}^{\bm n} \rho \mathbb{T}^{-\bm n}=\rho;$
          \item[(ii)] Finite correlation length (Definition~\ref{def:finitecorrlength});
         \item[(iii)] \label{cond:expsmallsubpurity}Exponentially small subsystem purity, $\tr(\rho_A^2)\leq e^{-\Omega(N_A)}$ for any hypercubical region $A$ with $N_A$ qubits.
     \end{enumerate}  Then, the average IPR in any  eigenbasis $\{\ket{j}\}_j$ of the translation operator is bounded
    \begin{equation}
    \label{eq:boundofIPR}
        \E_{\psi}\Big[\sum_{j=1}^{2^N} \abs{\braket{j}{\psi}}^4\Big]\leq O(N^{-1+\delta})
    \end{equation} for any $\delta>0$.
\end{theorem}
Theorem~\ref{th:boundonIPR} implies that a high-temperature Gibbs ensemble of shallow complexity states is an EDGE, taking  $\rho=g_\beta$ and $\ket{j}=\ket{E_j}$. Property (i) is satisfied as long as the Hamiltonian is TI, by Lemma~\ref{lemma:hightempgibbshavefinitecorr} property (ii) is satisfied if $\beta<\beta_*^{(\mathrm{corr})}$, and by Proposition~\ref{prop:bound_on_local_purity} property (iii) is satisfied if
$\beta<\beta_*^{(\mathrm{sep})}$. 

The proof of Theorem~\ref{th:boundonIPR} below generalizes that of Proposition~\maintextref{2} of the main text. The key idea to recall is that one may upper bound the average IPR by separately upper bounding (1) the overlap of the TI eigenstates with the aperiodic computational-basis states and (2) the probability of a bitstring being periodic. In generalizing this idea to arbitrary shallow complexity states, we face one immediate obstacle:  For an aperiodic bitstring $b\in \{0,1\}^{N}$ the state $\ket{b}$ is orthogonal to all of its translates,  $\bra{b}\mathbb{T}^{n}\ket{b}=0$ for all $0<n<N$. We relied on this orthogonality in deducing that $\abs{\braket{b}{j}}^2\leq 1/N$ for aperiodic $b$. However, for a general state  $\ket{\psi}$, it could be that $\bra{\psi}\mathbb{T}^{n}\ket{\psi}\neq 0$ even if $\ket{\psi}=\ket{\psi_1}\ket{\psi_2}\cdots \ket{\psi_N}$ is a product state and the sequence $\{\ket{\psi_{ 1}},\ket{\psi_{ 2}},\dots,\ket{\psi_{ N}}\}$ is aperiodic (e.g. $\ket{\psi}=\ket{+0000\cdots}$). To circumvent this problem, we establish a robust notion of {\it approximate} periodicity. We split the expectation value of the IPR into  an average over approximately periodic states and the rest, and prove that (1) the IPR of states which are not approximately periodic is upper bounded by $\sim 1/N$ and (2) the approximately periodic states are atypical, i.e., they are associated with vanishingly small probabilities. 

\subsection{Aperiodic states have low IPR}
 
 \begin{definition}(Approximately periodic state). An arbitrary state $\ket{\psi}$ for $N$ qubits arranged on the $D$-dimensional lattice $\Lambda$ is \emph{$r$-approximately periodic} if there exists $\bm n \in \Lambda$ with $\bm n \neq \bm 0$ such that $|\expval{  \mathbb{T}^ {\bm n}}{\psi}|^2\geq 2^{-r}$. The vector $\bm n$ is called a period of $\ket{\psi}$.\end{definition}

For now, $r$ is arbitrary. We will select an appropriate scaling $r(N)$ at the end of the proof. We begin by generalizing our decomposition of the average IPR from Proposition~\maintextref{2} for approximately periodic states. We state the result in a general form, which might be useful outside of our current context.
\begin{lemma}[Probabilistic bound on average IPR from symmetry]
\label{lemm:probabilisticboundonIPR}
    Consider a set of symmetry operators represented by mutually commuting unitaries $\{U_1, \dots,U_{M}\}$.  Let $P$ be the set of approximately symmetric states, containing all $\psi$ such that ${\abs{\expval{U_n^\dagger U_m}{\psi}}\geq \varepsilon}$ for some $n\neq m$. For any basis  $\{\ket{\alpha}\}_{\alpha}$ of simultaneous eigenstates of all $U_n$, the average IPR over  any ensemble of states $\mathcal{E}$  can be upper bounded as
    \begin{equation}
            \E_{\psi\in \mathcal{E}}\Big[\sum_\alpha\abs{\braket{\alpha}{\psi}}^4\Big]\leq \frac{1}{M}+\varepsilon+\mathbb{P}_{\psi\in \mathcal{E}}[\psi \in P],
    \end{equation}
    where  $\mathbb{P}_{\psi\in \mathcal{E}}[\psi \in P]$ denotes the probability that a state drawn from the ensemble $\mathcal{E}$ is  approximately symmetric.
\end{lemma}
\begin{proof}
To begin, we split the expectation value conditioning on $P$,
\begin{align}
\label{eq:boundbound}
    \E_{\psi\in \mathcal{E}}\left[\sum_\alpha\abs{\braket{\alpha}{\psi}}^4\right]&=\E_{\psi\in  \mathcal{E}}\Big[\sum_\alpha\abs{\braket{\alpha}{\psi}}^4 \Big| \psi \not \in P\Big]\mathbb{P}_{\psi\in  \mathcal{E}}[\psi \not \in P]+\E_{\psi\in  \mathcal{E}}\Big[\sum_\alpha\abs{\braket{\alpha}{\psi}}^4 \Big| \psi \in P\Big]\mathbb{P}_{\psi\in  \mathcal{E}}[\psi \in P]\\&
    \leq\E_{\psi\in  \mathcal{E}}\Big[\sum_\alpha\abs{\braket{\alpha}{\psi}}^4 \Big| \psi \not \in P\Big]+\mathbb{P}_{\psi\in  \mathcal{E}}[\psi \in P],
\end{align}
where the notation $\E[\, \cdots\,| \, \cdots \,]$ denotes a conditional expectation value.
Now we bound the first term, which is the expectation value conditioned on $\psi\not \in P$, i.e., $\psi$ not being approximately symmetric. We denote $\ket{\psi_n}=U_n\ket{\psi}$, which are approximately orthogonal states if $\psi \not \in P$, $\abs{\braket{\psi_n}{\psi_m}}< \varepsilon$ for $n\neq m$ . Moreover, observe that $\abs{\braket{\alpha}{\psi_n}}=\abs{\bra{\alpha}{U_n}\ket{\psi}}=\abs{\braket{\alpha}{\psi}}$, thus \begin{align}\label{eq:boundonoverlapsbeforenorms}
        M\abs{\braket{\alpha}{\psi}}^2=\sum_{n=0}^{M-1}\braket{\alpha}{\psi_n}\braket{\psi_n}{\alpha}\leq\expval{\big( \tilde \Pi - \Pi \big)}{\alpha} +1,
    \end{align}
where $\Pi$ is the orthogonal projector into the subspace spanned by $\{\ket{\psi_1},\dots,\ket{\psi_M}\}$ and  $\tilde{\Pi} \coloneqq \sum_n \dyad{\psi_n} $. We now show that these two operators are close to one another for $\psi \not \in P$.  Specifically, 
\begin{equation}
\label{eq:boundonnorms}
    \expval{\big( \tilde \Pi - \Pi \big)}{\alpha}^2\leq \|\tilde{\Pi}-\Pi\|^2_{\mathrm{op}}\leq \|{\tilde{\Pi}-\Pi}\|_{\mathrm{F}}^2= \mathrm{tr}\big( (\tilde{\Pi}-\Pi)^2\big)\leq \mathrm{tr}\big(\tilde{\Pi}^2)-M,
\end{equation}
where $\norm{A}_{\mathrm{F}}=\sqrt{\tr(AA^\dagger)}$ is the Frobenius norm, which always upper bounds the operator norm. We may compute $\mathrm{tr}\big(\tilde{\Pi}^2)=\sum_{n,m}\abs{\braket{\psi_n}{\psi_m}}^2\leq M(M-1)\varepsilon^2 +M\leq M^2\varepsilon^2+M$, which inserted into Eq.~\eqref{eq:boundonnorms} gives 
$\expval{\big( \tilde \Pi - \Pi \big)}{\alpha}\leq M\varepsilon$. Thus, from Eq.~\eqref{eq:boundonoverlapsbeforenorms} we obtain 
$\abs{\braket{\alpha}{\psi}}^2\leq 1/M+\varepsilon $ for each $\alpha$. Summing over all $\alpha$, we can bound the IPR of $\psi\not\in P$ as $\sum_\alpha \abs{\braket{\alpha}{\psi}}^4\leq \sum_\alpha \abs{\braket{\alpha}{\psi}}^2(1/M+\varepsilon)= 1/M+\varepsilon $, where we used the normalization $\sum_{\alpha}\abs{\braket{\alpha}{\psi}}^2=1$. Averaging over  $\psi\not\in P$, we obtain that
    \begin{equation*}
         \E_{\psi\in  \mathcal{E}}\Big[\sum_\alpha\abs{\braket{\alpha}{\psi}}^4 \Big| \psi \not \in P\Big] \leq \frac{1}{M}+\varepsilon. \qedhere
    \end{equation*}
\end{proof}

Lemma~\ref{lemm:probabilisticboundonIPR} provides a meaningful bound for the average IPR whenever $P$ is atypical in the ensemble $\mathcal{E}$ and $\varepsilon$ is small. We apply Lemma~\ref{lemm:probabilisticboundonIPR} with the unitaries  $U_n$ taken as certain translations on the lattice, which share a basis of eigenstates with the TI Hamiltonian. We select $P$ to be the set of $r$\nobreakdash-approximately periodic states with periods larger than the translations and show that $\mathbb{P}_{\psi\in  \mathcal{E}}\left [\psi \in P \right]$ is exponentially small.
 
Let $\delta>0$  (smaller than $1$). This is the same arbitrary $\delta$ that appears in the statement of Theorem~\ref{th:boundonIPR}, which will remain fixed henceforth. We consider lattice sites $\bm x_1,\bm x_2,\dots, \bm x_M\in \Lambda$ pairwise separated as $\norm{{\bm x_n}-{\bm x_m}}_1\geq L^{\delta}$ for all $n\neq m$. We can select at least $$M=\left(\left\lfloor\frac{L}{\lceil L^{\delta}\rceil}\right\rfloor -1\right)^D\geq \Omega(N^{1-\delta})$$ such sites (think of $\{\bm x_i\}_{i=1}^M$ as approximately forming a coarser sublattice). We take $U_n=\mathbb{T}^{\bm x_n}$, so that  $\abs{\expval{U_{m}^\dagger U_{n}}{\psi}}\geq  2^{-r/2}$ for $n\neq m$ implies that $\psi$ is $r$-approximately periodic with a period $\norm{\bm n}_1=\norm{\bm x_n -\bm x_m}_1\geq L^\delta$, i.e., 
$$P\coloneqq\left\{\psi\,\middle | \, \exists {\bm n} \in \Lambda \colon \norm{\bm n}_1\geq L^{\delta}, |\expval{  \mathbb{T}^ {\bm n}}{\psi}|^2\geq 2^{-r}\right\}.$$ 
 By Lemma~\ref{lemm:probabilisticboundonIPR} we obtain 
\begin{align}
\label{eq:IPRboundintermediate}
    \E_{\psi\in  \mathcal{E}}\Big[\sum_j\abs{\braket{j}{\psi}}^4\Big]\leq O\left(\frac{1}{N^{1-\delta}}\right)+{{2}^{-r/2}}+\mathbb{P}_{\psi\in  \mathcal{E}}\left [\psi \in P \right].
\end{align}
The reason to pick increasing periods $\norm{\bm n}_1\geq L^{\delta}$ is that this will guarantee that $\mathbb{P}_{\psi\in \mathcal{E}}[\psi \in P]$ is exponentially small, as we prove next.

\subsection{Approximately periodic states are atypical}
Intuitively, periodic states with large periods will be long-range correlated, so they cannot have a significant population in an ensemble whose average state has finite correlation length. Specifically, we bound $\mathbb{P}_{\psi\in  \mathcal{E}}\left [\psi \in P \right]$ by separately bounding the probability of a state having fixed period $\bm n$, i.e. of being in
    \begin{equation}
        P_{\bm n}=\left\{\psi\,\middle |\, \abs{\expval{\mathbb{T}^{\bm n }}{\psi}}^2 \geq 2^{-r}\right\},
    \end{equation}
and then taking the union bound over $P=\bigcup_{\norm{\bm n}_1\geq L^{\delta}} P_{\bm n}$.

\begin{prop}[Bound on probability of approximately periodic states]
\label{prop:boundonprobabilityofperiodicityarbdim}
Consider an ensemble of pure states $\mathcal{E}$ with depth complexity $\mathcal{C}$, whose mean state $\rho=\E_{\psi \in\mathcal{E}}[\dyad{\psi}]$ is translation invariant and has finite correlation length~$\xi$. Let ${\bm n }\in \Lambda$ and consider a hypercubical region $A$ with side length $\ell$  and $N_A=\ell^D$ such that $\norm{\bm n }_1 - D\ell> 2\mathcal{C}$. Then,

\begin{equation}
\label{eq:limitonP}
    \Pro\limits_{\psi \in\mathcal{E}}\left[\psi\in P_{\bm n}\right]\leq 2^{r+2\mathcal{C}D\ell^{D-1}}\Big(2^{(\ell^D)}f_{\mathrm{corr}}(2D\ell^{D-1}) e^{-(\norm{\bm n }_1-D\ell)/\xi}+\tr(\rho_A^2)\Big),
\end{equation}
where $f_\mathrm{corr}$ is the polynomial function from  Definition~\ref{def:finitecorrlength}.
\end{prop}
\begin{proof}

    Let $O_A=\bigotimes_{\bm i\in A} \sigma_{\bm i}$ be an arbitrary product of Pauli matrices $\sigma_{\bm i}\in \{X,Y,Z,\mathds{1}\}$ defined over $A$ (a so-called {\it Pauli string}). Let $A+\bm n$ denote the translation of the region $A$ over the vector ${\bm n }$ and $O_{A+\bm n}=\mathds{T}^{\bm n} O_A \mathds{T}^{-\bm n} $. Note that the distance between $A$ and $A+\bm n$ is at least $\norm{\bm n}_1 -D\ell$. Then, because $\rho$ has finite correlation length $\xi$ and the number of qubits in the boundary of $A$ is $\abs{\partial A}=2D\ell^{D-1}$,
     \begin{align}
          f_{\mathrm{corr}}(2D\ell^{D-1})e^{-(\norm{\bm n }_1-D\ell)/\xi }&\geq  \tr(\rho O_A\otimes  O_{A+\bm n}) - \tr(\rho_{A} O_A)\tr(\rho_{A+\bm n} O_{A+\bm n}) \\&= \tr(\bigg(\E_{\psi\in  \mathcal{E}} [\psi_{ A}  \otimes \psi_{A+\bm n}] - \rho_{A}\otimes \rho_{A+\bm n}\bigg)O_A\otimes O_{A+\bm n}),
    \end{align}
    where we used the short-range correlation of $\psi$, Lemma~\ref{lemm:propsofconstdepthcomplexity}~{\it (ii)}, which guarantees that $\tr(\psi O_A\otimes O_{A+\bm n})=\tr(\psi O_A)\tr(\psi O_{A+\bm n})$ whenever $d(A,A+\bm n)> 2\mathcal{C}$, which is the case here.
    Summing over all $4^{(\ell^D)}$ possible Pauli strings over $A$, and using that $\sum_{O_{A}} {O}_A\otimes  O_{A+\bm n} = 2^{(\ell^D)}S_{A,A+\bm n}$, where $S_{A,A+\bm n}$ is the swap operator between regions $A$ and $A+\bm n$, we get 
     \begin{align}
     \label{eq:corrinequalitya}
         2^{(\ell^D)}  f_{\mathrm{corr}}(2D\ell^{D-1})e^{-(\norm{\bm n }_1-D\ell)/\xi }&\geq  \tr(\bigg(\E_{\psi\in  \mathcal{E}} [\psi_{ A}  \otimes \psi_{A+\bm n}]- \rho_{A}\otimes \rho_{A+\bm n} \bigg)S_{A,A+\bm n})\\
         \nonumber
        &=\E_{\psi\in  \mathcal{E}}[\tr(\psi_{ A}   \psi_{A+\bm n})] - \tr(\rho_{A}^2), 
        \end{align}
where we used that $\tr(S (W\otimes V)) =\tr(WV)$ and  the translation invariance $ \rho_A=\rho_{A+\bm n}$.
Now, we leverage Lemma \ref{lemm:propsofconstdepthcomplexity}~{\it (i)} to bound $\tr(\psi_A \psi_{A+\bm n})$ in terms of $\abs{\langle \psi|\mathbb{T}^{\bm n}|\psi \rangle}^2$. Because $\abs{\partial_{\mathcal{C}}A}=\abs{\partial_{\mathcal{C}}(A+\bm n)} \leq 2\mathcal{C}D\ell^{D-1}$, both $\psi_A$ and $\psi_{A+\bm n}$ have rank at most $R=2^{2\mathcal{C} D \ell^{D-1}}$, and the inequality between the trace norm and the Frobenius norm gives us 
    $$\tr(\psi_A \psi_{A+\bm n})=\norm{\sqrt{\psi_A}\sqrt{\psi_{A+\bm n}}}_{\mathrm{F}}^2\geq\frac{1}{R}\norm{\sqrt{\psi_{A}}\sqrt{\psi_{A+\bm n}}}_{\mathrm{tr}}^2\geq\frac{1}{R}\abs{\langle \psi|\mathbb{T}^{\bm n}|\psi \rangle}^2.$$
    The last inequality follows from the fact the fidelity of reduced density matrices cannot be smaller than the global fidelity (this follows from Uhlmann's theorem) \cite{BookNielsen2010}. Finally, if $\psi\in P_{\bm n}$, then $\abs{\langle \psi|\mathbb{T}^{\bm n}|\psi \rangle}^2\geq 2^{-r}$, so
        \begin{equation}       
        \E_{\psi\in  \mathcal{E}}[\tr(\psi_{ A}   \psi_{A+\bm n})]\geq \frac{1}{R}\E_{\psi\in  \mathcal{E}}[\abs{\langle\psi |\mathbb{T}^{\bm n}|\psi \rangle}^2 ] \geq 2^{-(r+2\mathcal{C}D\ell^{D-1})}\Pro\limits_{\psi \in\mathcal{E}}\left[\psi \in P_{\bm n}\right]. 
\label{eq:inequality1proofcorrtranslates}
        \end{equation}
        By inserting Eq.~\eqref{eq:inequality1proofcorrtranslates} into Eq.~\eqref{eq:corrinequalitya} and solving for $\mathbb{P}_{\psi\in \mathcal{E}}[\psi\in P_{\bm n}]$, we complete the proof.
\end{proof}
Performing a union bound over all $\norm{\bm n }_1\geq L^{\delta}$ in Eq.~\eqref{eq:limitonP}, we obtain
\begin{align}
\label{eq:boundonnprobP}
    \mathbb{P}_{\psi\in  \mathcal{E}}\left [\psi \in P \right]&\leq N 2^{r+2\mathcal{C}D\ell^{D-1}}  \Big(2^{(\ell^D)} f_{\mathrm{corr}}(2D\ell^{D-1}) e^{-(L^{\delta}-D\ell)/\xi}+\tr(\rho_{A}^2)\Big).
\end{align}
 The term $\tr(\rho_A^2)$ is the subsystem purity of $\rho$, which by condition \ref{cond:expsmallsubpurity} is exponentially small in $N_A=\ell^D$, i.e.,  $\tr(\rho_A^2)\leq e^{-\Omega(\ell^D)}$.
 Thus, from Eq.~\eqref{eq:boundonnprobP} we get the asymptotic bound on the probability of a state being $r$-approximately periodic:
\begin{equation}
\label{eq:asymptoticboundonprob}
    \mathbb{P}_{\psi\in  \mathcal{E}}\left [\psi \in P \right]\leq N 2^{r+O(\mathcal{C}\ell^{D-1})}\Big(\mathrm{poly}(\ell)e^{O(\ell^D)}e^{- \Omega(L^{\delta})}+ e^{-\Omega(\ell^D)}\Big) ,
\end{equation}
where $\text{poly}(\ell)$ denotes a polynomial on $\ell$. 

Inserting Eq.~\eqref{eq:asymptoticboundonprob} into Eq.~\eqref{eq:IPRboundintermediate}  gives the bound on the IPR,
\begin{align}
\label{eq:asymptoticboundonIPR}
    \E_{\psi\in  \mathcal{E}}\bigg[\sum_j\abs{\braket{j}{\psi}}^4\bigg]\leq O\Big(\frac{1}{ N^{1-\delta}}\Big)+ \underbrace{N \mathrm{poly}(\ell)e^{O(\ell^D)- \Omega(L^{\delta})+ O(\mathcal{C}\ell^{D-1}) + O(r)}}_{(\star)}+ \underbrace{N  e^{O(r) + O(\mathcal{C}\ell^{D-1})-\Omega(\ell^D)}}_{(\star\star)} + \underbrace{e^{-\Omega(r)}}_{(\star\star\star)} .
\end{align}
We are free to select how $\ell$ and $r$ scale with $L=N^{1/D}$. We select the scaling so the first term on the right-hand side of Eq.~\eqref{eq:asymptoticboundonIPR} dominates over the other three terms. Specifically, we pick
\begin{align}
 \ell=\Theta(L^{a }) &&\text{and}&& r=\Theta(L^{b})
\end{align}
so
$(\star\star\star)= e^{-\Omega(L^b)}$. Furthermore, because the complexity grows subpolynomially, $\mathcal{C}\leq O(L^{\nu})$ for any $\nu>0$,  we have $\mathcal{C}\ell^{D-1}\leq O(L^{(D-1)a+\nu})$. Thus, as long as $0<b<aD<\delta$ and $\nu<a$, we have $(D-1)a+\nu< aD<\delta$ and consequently $(\star)\leq e^{-\Omega(L^\delta)}\leq e^{-\Omega(L^{b})}$ and $(\star\star)\leq e^{-\Omega(L^{aD})}\leq e^{-\Omega(L^{b})}$.

All in all, the last three terms on the right-hand side of Eq.~\eqref{eq:asymptoticboundonIPR} are upper bounded by $e^{-\Omega(L^{b})}$, over which the first term dominates. Then we may conclude, 
\begin{align}
    \E_{\psi\in  \mathcal{E}}\bigg[\sum_j\abs{\braket{j}{\psi}}^4\bigg]\leq O\left(\frac{1}{N^{1-\delta}}\right),
\end{align}
which is the statement of Theorem~\ref{th:boundonIPR}.

\section{Thermalization to generalized Gibbs states}
\label{sec:generalized}
In this section we extend our proof of quantum thermalization beyond the canonical ensemble $g_\beta\propto e^{-\beta H}$. We consider systems with additional conserved quantities where the general form of the equilibrium state has additional structure, for example a grand canonical ensemble or a generalized Gibbs ensemble \cite{Vidmar2016}. Specifically, we consider a set of local  conserved quantities $Q_i$, meaning that $[H,Q_i]=0$ and $Q_i$ can can be decomposed as a sum of observables with bounded operator norm and each acting on a geometrically local region.   The generalized Gibbs state (GGS) is parametrized by a vector $\bm \lambda=(\lambda_1,\lambda_2,\dots,\lambda_m)$ of  {\it generalized chemical potentials} (GCP) as
\begin{equation}
\label{eq:GGE}
    g_{\bm \lambda}=\frac{\exp(-\sum_{j=1}^m\lambda_j Q_j)}{\tr(\exp(-\sum_{j=1}^m\lambda_j Q_j))}.
\end{equation}
  For example if we have energy conservation $Q_1=H$  and one single additional $U(1)$ symmetry (e.g., $Q_2=\sum_{\bm j \in \Lambda} Z_{\bm j}$), the state $g_{(\lambda_1=\beta,\lambda_2=-\beta\mu)}$ describes a grand canonical ensemble with chemical potential $\mu$.
  If the system is integrable and thus supports an extensive number of conserved quantities, there may be a chemical potential associated to each of them.

To state our main result on thermalization to the GGS, we first need to generalize our notion of MEE.

\begin{definition}[Generalized maximally entropic  ensemble (G-MEE)]
    Given a vector of GCPs $\bm \lambda$, we say an ensemble of pure states $\mathcal{E}_{\bm \lambda}$ has GCPs $\bm \lambda$ if the average expectation value of the conserved quantities are equal to those of the GGS, \begin{equation}
        \forall j:\E_{\psi\in\mathcal{E}_{\bm \lambda}}[\expval{{Q_j}}{\psi}]=\tr(g_{\bm \lambda}Q_j).
    \end{equation} Furthermore, an ensemble 
 $\mathcal{E}^*_{\bm \lambda,\mathcal{C}}$ with GCPs $\bm \lambda$ and complexity $\mathcal{C}$ is a G-MEE if it maximizes the von Neumann entropy of its average state among all ensembles with the same GCPs and complexity,
    \begin{equation}
\mathcal{E}^*_{\bm{\lambda},\mathcal{C}}\in \mathop{\mathrm{argmax}}_{\substack{\mathcal{E}_{\bm{\lambda},\mathcal{C}}}} \,S(\rho_{\bm{\lambda},\mathcal{C}}).
\end{equation}
\end{definition}
We establish a proof of quantum thermalization to GGSs.  The GCPs play the role of inverse temperature $\beta$ in Theorem~1, so the  statement about thermalization at high temperature (or equivalently small $\beta$) is here promoted to a statement about small GCPs, in the sense that each entry of $\bm \lambda$ is below a threshold constant. 
\begin{theorem}[Quantum thermalization to generalized Gibbs states]
\label{th:G-main}
 Consider any translation invariant, geometrically local Hamiltonian $H$ for $N$ qubits in $D$-dimensional lattice with nondegenerate spectral gaps.
 Let $Q_1,\dots,Q_m$ be a constant number of local translation invariant conserved quantities $[Q_i,H]=0$. 
 There exists a threshold constant $\lambda_*>0$ such that for any $\bm \lambda=(\lambda_1,\dots,\lambda_m)$ with generalized chemical potentials bounded as $\abs{\lambda_i}\leq \lambda_*$, any G-MEE with shallow complexity $\mathcal{E}_{\bm \lambda,\mathcal{C}}$  
 thermalizes, 
 \begin{equation}
     \E_{\psi \in \mathcal{E}_{{\bm \lambda}, \mathcal{C}}}\E_t[\norm{\psi(t)-g_{\bm \lambda}}_A]\leq O\big(N^{-{1/2}+\delta}\big)
 \end{equation}
for any geometrically local region $A$ and 
for any $\delta>0$.
\end{theorem}
Before moving on to explain the proof, two remarks are in order. First, note that Theorem~\ref{th:G-main} considers a constant number of conserved quantities $m$ which should be independent of $N$. Nevertheless, Theorem~\ref{th:G-main} still applies to the dynamics generated by a Hamiltonian with more local conserved quantities, such as Bethe ansatz integrable systems. We note, however, that in such systems one can only impose a constant number of nonzero GCPs for the ensemble of initial states. As we will see below, the threshold GCP $\lambda_*$ depends on $m$ and on the choice of $Q_1,\dots, Q_m$.

The second remark is that, like Theorem~1, Theorem~\ref{th:G-main} requires the nondegenerate gap condition, which also requires no degeneracies in the spectrum. In particular, we require the Hamiltonian to be free of any nonabelian symmetries (which necessarily imply degeneracy). This has two consequences: first, all the conserved quantities must commute pairwise $[Q_i,Q_j]=0$, and secondly, the system cannot possess a spatial inversion symmetry (which does not commute with the translation operator). The latter consequence somewhat hinders the applicability of Theorem~\ref{th:G-main} to certain interacting integrable spin chains, such as the Heisenberg or XXZ/XYZ models  which  usually posses a spatial inversion symmetry. Nevertheless we note that one can  construct integrable spin chains without inversion symmetry\footnote{We thank Nicholas O'Dea for pointing this out to us.}, to which Theorem~\ref{th:G-main} applies. 
It is an interesting and important direction to extend our results to systems with nonabelian symmetries. 

The proof of Theorem~\ref{th:G-main} follows the same path as Theorem~1. We prove a generalized EDGE (G-EDGE) Theorem, and then we show that any shallow-complexity G-MEE at low GCP is a G-EDGE. 
\subsection{Properties of the generalized Gibbs state}

We begin by proving some properties of the GGS. 
\begin{prop}[Properties of generalized Gibbs states at low chemical potential]
   \label{proposition:propsofGGS}
    Let $Q_1,Q_2,\dots Q_m$ be local observables, where $m$ is a constant. Let $\bm \lambda=(\lambda_1,\dots,\lambda_m)$ be any GCP vector.
    \begin{enumerate}
     \item[(a)] [Maximum entropy principle]. Among all states $\rho$ satisfying the constraints $\tr(\rho Q_i) = W^{(i)}_{\bm \lambda}$, where $W^{(i)}_{\bm \lambda}\coloneqq\tr(g_{\bm \lambda}Q_i)$, the GGS $g_{\bm \lambda}$ uniquely maximizes the von Neumann entropy 
        \begin{equation}
            g_{\bm \lambda}=\mathop{\mathrm{argmax}}_{\rho:\tr(\rho Q_i)=W^{(i)}_{\bm \lambda}} S(\rho).
        \end{equation}
        \item[(b)] [Finite correlation length]. There exists a positive constant $\lambda_*^{(\mathrm{corr})}$ such that $g_{\bm \lambda}$ has finite correlation length  whenever  $\max_i\abs{\lambda_{i}} \leq \lambda_*^{(\mathrm{corr})}$.
        \item[(c)] [Separability]. There exists a positive constant $\lambda_*^{(\mathrm{sep})}$ such that $g_{\bm \lambda}$ is separable  whenever  $\max_i\abs{\lambda_{i}} \leq \lambda_*^{(\mathrm{sep})}$.
        \item[(d)] [Small subsystem purity]. If $\max_i\abs{\lambda_{i}} \leq \lambda_*^{(\mathrm{corr})}$, then $g_{\bm \lambda}$ has an exponentially small subsystem purity, $\tr((g_{\bm \lambda,A})^2)\leq e^{-\Omega(N_A)}$ for any region $A$ with $N_A$ qubits.

    \end{enumerate} 
    \begin{proof}
         Property (a) follows from a standard argument: First, we compute the relative entropy \begin{equation}
            D(\rho || g_{\bm \lambda})=-S(\rho)+\sum_i\lambda_i\tr(Q_i \rho) + \log(\tr(e^{-\sum_i\lambda_i Q_i})).
         \end{equation}
         Second, we impose the constraints $\tr(\rho Q_i)=W^{(i)}_{\bm \lambda}$, and thus we see than maximizing $S(\rho)$ is equivalent to minimizing the relative entropy, whose unique minimum $D(\rho || g_{\bm \lambda})=0$ is attained by  $\rho=g_{\bm \lambda}$.
         
         For properties (b), (c), and (d) the idea is simply to write $g_{\bm \lambda}$ as a regular Gibbs state for some local Hamiltonian $\widetilde{H}$ and inverse temperature $\widetilde{\beta}$, and leverage the corresponding properties that we already established for the Gibbs state at high temperature. Specifically, we define \begin{equation}\label{eq:hamtilde}\widetilde{H}_{\bm \lambda}=\frac{1}{\max_i \abs{\lambda_{i}}}\sum_{i=1}^m\lambda_iQ_i 
\end{equation} 
and $\widetilde \beta_{\bm \lambda}=\max_i \abs{\lambda_{i}}$ so that immediately $\widetilde g_{\widetilde \beta_{\bm \lambda}}\coloneqq \mathrm{exp}(-\widetilde{\beta}_{\bm \lambda} \widetilde{H}_{\bm \lambda})/\mathrm{tr}(\mathrm{exp}(-\widetilde{\beta}_{\bm \lambda} \widetilde{H}_{\bm \lambda}))=g_{\bm \lambda}$.
Furthermore, the Hamiltonian $\widetilde{H}_{\bm \lambda}$ is local, because $m$ is a constant and each $Q_i$ is local.  

We begin by proving separability.  If we apply Lemma~\ref{lemma:hightempgibbshavefinitecorr} to $\widetilde{H}_{\bm \lambda}$ we obtain a constant inverse temperature  $\widetilde{\beta}^{(\mathrm{sep})}_*>0$ such that for any $\widetilde\beta\leq\widetilde{\beta}^{(\mathrm{sep})}_*$, the state $\widetilde{g}_{\widetilde{\beta}}$ is separable. The key point to note is that the constant $\widetilde\beta_*^{(\mathrm{sep})}$ is actually independent of $\bm \lambda$ (for generic choices such that $\lambda_i\neq 0$, which we assume without loss of generality). This is because the separability threshold $\widetilde\beta_*^{(\mathrm{sep})}$ only depends on the locality properties of the Hamiltonian $\widetilde H_{\bm \lambda}$, and is independent of the specific relative weights of its terms. Thus, we can define $\lambda_*^{(\mathrm{sep})}\coloneqq \widetilde\beta_*^{(\mathrm{sep})}$. Then, for any $\bm \lambda$ such that $\widetilde{\beta}_{\bm \lambda}\coloneqq \max_i \abs{\lambda_i} \leq \lambda_*^{(\mathrm{sep})}$,
we have $\widetilde{\beta}_{\bm \lambda}\leq \widetilde{\beta}^{(\mathrm{sep})}_*$, so  $g_{\bm \lambda}={\widetilde g}_{\widetilde\beta_{\bm \lambda}}$ is separable.

We note that, in general, $\widetilde{\beta}^{(\mathrm{sep})}_*$ depends on the locality properties of $\widetilde{H}_{\bm \lambda}$ such as the number of terms acting on a certain qubit. For our purposes, this translates to that the associated $\lambda_*^{(\mathrm{sep})}$ depends on the number of conserved quantities $m$.

The argument for finite correlation length is the same, by now applying Lemma~\ref{lemma:hightempgibbshavefinitecorr} to $\widetilde{H}_{\bm \lambda}$. The exponentially small subsystem purity follows from Proposition~\ref{prop:bound_on_local_purity}, applied to $g_{\bm \lambda}={\widetilde g}_{\widetilde\beta}$.
    \end{proof}
\end{prop}

\subsection{Generalized EDGE theorem}
\label{subsec:GEDGE}
Now, we generalize the EDGE theorem. For this we need to generalize the concept of EDGEs.
\label{def:genEDGE}
\begin{definition}[Generalized energy dispersed Gibbs ensemble] An ensemble of states $\mathcal{E}_{\bm \lambda}$ is a generalized EDGE (G-EDGE) if it satisfies the following two conditions:
\begin{enumerate}
    \item[(a)] [Generalized Gibbs ensemble]. The average of the ensemble is close to $g_{\bm \lambda}$,
    \begin{equation}
        \Big\|\E_{\psi\in \mathcal{E}_{\bm \lambda}}[\psi]- g_{\bm \lambda}\Big \|\leq2^{-N}\varepsilon_{\mathrm{GE}}(N).
    \end{equation} 
    \item[(b)][Energy dispersion]. The average IPR in the energy eigenbasis is upper bounded as
    \begin{equation}
         \E_{\psi\in \mathcal{E}_{\bm \lambda}}\Big[\sum_{j=1}^{2^N} \abs{\braket{E_j}{\psi}}^4\Big]\leq\varepsilon_{\mathrm{ED}}(N).
    \end{equation}
\end{enumerate}
\end{definition}
We can prove a G-EDGE theorem:
\begin{theorem}[Generalized EDGE]
\label{th:G-EDGE}
         Let $H$ be a TI, geometrically local Hamiltonian with nondegenerate spectral gaps on a $D$-dimensional periodic lattice with $N$ qubits and $\mathcal{E}_{\bm \lambda}$ be an G-EDGE. Assume that the generalized Gibbs state $g_{\bm \lambda}$ has finite correlation length. Then $       \E_{\psi\in \mathcal{E}_{\bm \lambda}} \E_{t\geq 0} \left[\norm{g_{\bm \lambda}-\psi(t)}_A \right]$ is upper bounded by the right-hand side of Eq~\eqref{eq:boundav}. At sufficiently small chemical potential $\max_i\abs{\lambda_i}<\lambda_*^{(\mathrm{con})}$,  we can take any $\gamma<1/2$.\footnote{Note that a finite-time statement analogous to Theorem~\ref{th:finite-time-EDGE} can also be proven.}
         \begin{proof}
             The proof is the same as the EDGE Theorem~\ref{th:EDGE}, with the only difference that the equivalence of temporal and thermal ensembles now reads 
              \begin{equation}
                  \E_{\psi\in  \mathcal{E}_{\bm \lambda}} [\norm{\rho_\infty-g_{\bm \lambda}}_A]=O(N^{-\gamma})+2\varepsilon_{\mathrm{GE}}(N).
              \end{equation}
              We prove this in the same manner as Proposition~\ref{prop:EDGEeqofens}, simply replacing $\beta$ by $\bm \lambda$. Note that Corollary~\ref{cor:similarformofweakETH} also applies to $g_{\bm \lambda}$ because this state is also diagonal in the eigenbasis of $H$, as $[Q_i,H]=0$.  Furthermore, to obtain the improved scaling $\gamma<1/2$, we simply write $g_{\bm \lambda}$ as a high-temperature Gibbs state (as in the proof of \ref{proposition:propsofGGS}), and the constant $\lambda_*^{(\mathrm{con})}$ is obtained in the same manner as the constants in Proposition~\ref{proposition:propsofGGS}, from the threshold temperature  $\beta_*^{(\mathrm{con})}$ of Proposition~\ref{prop:weakETH}. 
         \end{proof}

\end{theorem}

\subsection{Shallow-complexity G-MEEs are G-EDGEs at low generalized chemical potential}
Finally, we prove that any G-MEE of shallow-complexity is an G-EDGE. We first have an analogous result to Lemma~\ref{lemma:HT-MEE-GE}, which establishes property (a) of a G-EDGE with $\varepsilon_{\mathrm{GE}}=0$.
\begin{lemma}[G-MEE average to generalized Gibbs states at low generalized chemical potential.]
\label{lemm:G-MEEareGSS}
Let $Q_1,Q_2,\dots Q_m$ be local observables, where $m$ is a constant. For any complexity function $\mathcal{C}(N)$, there exists a positive constant $\lambda_*^{(\mathcal{C})}\geq \lambda_*^{(\mathrm{sep})}$ such that for any chemical potential vector $\bm \lambda=(\lambda_1,\dots,\lambda_m)$ with $\max_{i}\abs{\lambda_{i}} \leq \lambda_*^{(\mathcal{C})}$, any G-MEE $\mathcal{E}_{\bm\lambda,\mathcal{C}}$ averages to the generalized Gibbs state,
    \begin{equation}
        \E_{\psi\in \mathcal{E}_{\bm{\lambda},\mathcal{C}}}[\psi]=g_{\bm \lambda}.
    \end{equation}
\end{lemma}
\noindent The proof is the same as Lemma~\ref{lemma:HT-MEE-GE}, replacing $\beta$ with $\bm \lambda$ and leveraging the maximum entropy principle and separability at low GCPs 
proven in Proposition~\ref{proposition:propsofGGS}~(a) and~(c), respectively.

Property (b) of a G-EDGE follows simply from Theorem~3. The GGS at low $\bm \lambda$, i.e. with $\max_j|\lambda_j|\leq \lambda_*\coloneqq\min\{\lambda_*^{(\mathrm{sep})},\lambda_*^{(\mathrm{corr})},\lambda_*^{(\mathrm{con})}\}$ is TI whenever each $Q_i$ is TI, and has finite correlation length and an exponentially small subsystem purity by Proposition~\ref{proposition:propsofGGS}~(b) and (d), respectively. Thus, by Theorem 3, we obtain a bound on the average IPR $\varepsilon_{\mathrm{ED}}=O(N^{-1+\delta})$, which combined with Lemma~\ref{lemm:G-MEEareGSS} and the G-EDGE Theorem~\ref{th:G-EDGE} gives us Theorem~\ref{th:G-main}.
\let\oldaddcontentsline\addcontentsline
\renewcommand{\addcontentsline}[3]{}
\bibliography{references}
\let\addcontentsline\oldaddcontentsline